\documentclass[a4paper,11pt]{article}
\pdfoutput=1

\usepackage{afterpage}
\usepackage{comment}
\usepackage{subcaption}
\usepackage[english]{babel}
\usepackage[utf8x]{inputenc}
\usepackage{slashed}
\usepackage{amsmath}
\usepackage{tikz}
\usepackage{multicol}
\usepackage{multirow}
\usepackage{jheppub} 
\allowdisplaybreaks

\usepackage{tikz,xcolor,hyperref}

\definecolor{lime}{HTML}{A6CE39}
\DeclareRobustCommand{\orcidicon}{\hspace{-4pt}
	\begin{tikzpicture}
		\draw[lime, fill=lime] (0,0) 
		circle [radius=0.16] 
		node[white] {\hspace{0.1mm}{\fontfamily{qag}\selectfont \tiny ID}};
		\draw[white, fill=white] (-0.07,0.1) 
		circle [radius=0.01];
	\end{tikzpicture}
	\hspace{-3.2mm}
}

\foreach \x in {A, ..., Z}{\expandafter\xdef\csname 
	orcid\x\endcsname{\noexpand\href{https://orcid.org/\csname orcidauthor\x\endcsname}
		{\noexpand\orcidicon}}
}

\def\gs{\mathrel{
   \rlap{\raise 0.511ex \hbox{$>$}}{\lower 0.511ex \hbox{$\sim$}}}}
\def\ls{\mathrel{
   \rlap{\raise 0.511ex \hbox{$<$}}{\lower 0.511ex \hbox{$\sim$}}}}

\preprint{TIFR/TH/22-44}

\title{Sterile Neutrinos: Propagation in Matter and Sensitivity to Sterile Mass Ordering} 

\author[a]{Dibya S. Chattopadhyay\orcidA{}}
\author[b]{Moon Moon Devi\orcidB{}}
\author[a]{Amol Dighe\orcidC{}}
\author[c]{Debajyoti Dutta\orcidD{}}
\author[d]{Dipyaman Pramanik\orcidE{}}
\author[e]{Sushant K. Raut\orcidF{}}

\affiliation[a]{Tata Institute of Fundamental Research, Homi Bhabha Road, Colaba, Mumbai 400005, India}
\affiliation[b]{Department of Physics, Tezpur University, Assam, 784028, India}
\affiliation[c]{Assam Don Bosco University, Tapesia Campus, Sonapur, Assam, 782402 India}
\affiliation[d]{
Instituto de Física Gleb Wataghin - UNICAMP, 13083-859, Campinas, São Paulo, Brazil}
\affiliation[e]{Division of Sciences, Krea University, Sri City, India 517646}

\emailAdd{d.s.chattopadhyay@theory.tifr.res.in}
\emailAdd{devimm@tezu.ernet.in}
\emailAdd{amol@theory.tifr.res.in}
\emailAdd{debajyoti.dutta@dbuniversity.ac.in}
\emailAdd{dpramanik92@gmail.com}
\emailAdd{sushant.raut@krea.edu.in}

\abstract{
We analytically calculate the neutrino conversion probability $P_{\mu e}$ in the presence of sterile neutrinos, with exact dependence on $\Delta m^2_{41}$ and with matter effects explicitly included. Using perturbative expansion in small parameters, the terms involving the small mixing angles $\theta_{24}$ and $\theta_{34}$ can be separated out, with $\theta_{34}$ dependence only arising due to matter effects. 
We express $P_{\mu e}$ in terms of the quantities of the form $\sin(x)/x$, which helps in elucidating its dependence on matter effects and a wide range of $\Delta m^2_{41}$ values.
Our analytic expressions allow us to predict the effects of the sign of $\Delta m^2_{41}$ at a long baseline experiment like DUNE.
We numerically calculate the sensitivity of DUNE to the sterile mass ordering and find that this sensitivity can be significant in the range $|\Delta m^2_{41}| \sim (10^{-4} - 10^{-2})$ eV$^2$, for either mass ordering of active neutrinos.
The dependence of this sensitivity on the value of $\Delta m^2_{41}$ for all mass ordering combinations can be explained by investigating the resonance-like terms appearing due to the interplay between the sterile sector and matter effects.
}

\begin{document} 
\maketitle

\section{Introduction}
\label{sec:intro}
The phenomenon of neutrino oscillations, originating from different
masses of three active neutrinos and mixing among the three neutrino flavors, 
is now well-established, and explains all the data from solar, atmospheric, 
and reactor neutrinos quite well~\cite{SajjadAthar:2021prg,Workman:2022ynf}.
The magnitudes of mass-squared differences $\Delta m^2_{ij} \equiv m_i^2 - m_j^2$ between each pair of neutrino mass eigenstates, as well as the mixing
angles $\theta_{ij}$ that parameterize the neutrino mixing matrix, have been
measured to an accuracy of better than 10\%, and the sign of $\Delta m^2_{21}$
has been determined to be positive from the solar neutrino data \cite{Esteban:2020cvm, nufit,deSalas:2020pgw,Capozzi:2021fjo}. 
Two of the parameters controlling neutrino mixing and oscillations
that have not been determined so far are the mass ordering, i.e. the
sign of $\Delta m^2_{31}$, and the CP-violating phase $\delta_{13}$.
Many of the future experiments~\cite{Hyper-KamiokandeProto-:2015xww,ICAL:2015stm,JUNO:2015zny,DUNE:2020ypp,DUNE:2020jqi} have the measurement
of these two quantities as one of their primary aims.

While the data from the collider experiments~\cite{ALEPH:2005ab} have ruled out
the presence of more than three light neutrinos that undergo weak 
interactions, the possibility of one or more sterile neutrino species, 
which do not undergo weak interactions, remains.
The existence of a sterile neutrino is a crucial question that connects to our quest for a fundamental theory at the high scale, from which the standard model (SM) of particle physics would emerge as an effective theory.
Indeed, even the origin of the masses of active neutrinos themselves
needs the introduction of new physics at the high scale~\cite{doi:10.1142/5024,giunti2007fundamentals,King:2014nza,zuber2020neutrino,Dasgupta:2021ies}. Any light fermion in these 
theories that is a SM gauge singlet can mix with active neutrinos and 
play the role of a sterile neutrino.

Some short-baseline accelerator experiments have claimed observations 
that would need the presence of sterile neutrinos for their explanation~\cite{Athanassopoulos:1995iw,Aguilar:2001ty,Gariazzo:2018mwd,MiniBooNE:2018esg,PhysRevD.103.052002}.
There are also indications that the reactor neutrino data may be
accounted for better in the presence of sterile neutrinos~\cite{Mention:2011rk,Mueller:2011nm,Huber:2011wv,Gariazzo:2018mwd,Collin:2016aqd,Gariazzo:2017fdh}. However, the evidence is still inconclusive 
and in tension with other experiments~\cite{Esmaili:2013vza,Collin:2016aqd,Capozzi:2016vac,Gariazzo:2017fdh,Dentler:2018sju,MicroBooNE:2021tya,MicroBooNE:2022wdf}, as well as recent theoretical calculations related to reactor nuclear effects~\cite{Giunti:2021kab}.
The short-baseline experiments mentioned above need sterile neutrinos
with $\Delta m^2_{41} \sim 0.1-1$ eV$^2$ for explaining the data, where $m_4$ 
is the mass of the eigenstate with the largest sterile neutrino component.
Short-baseline Gallium radioactive source experiments like GALLEX~\cite{GALLEX:1994rym,GALLEX:1998kcz,Kaether:2010ag}, SAGE~\cite{SAGE:1998fvr,SAGE:1999uje,Abdurashitov:2005tb,SAGE:2009eeu} and BEST~\cite{Barinov:2021asz,Barinov:2022wfh} have measured electron neutrino disappearance levels that are much higher than expected. However, note that the sterile neutrino mixing angles needed to solve such anomalies are in tension~\cite{Giunti:2022btk} with the constraints from other experiments, and may need inclusion of more exotic new physics scenarios~\cite{Arguelles:2022bvt}.
Sterile neutrinos of keV masses have also been proposed as candidates for 
warm dark matter in theories like the $\nu$SM~\cite{Asaka:2005an,Asaka:2005pn,Boyarsky:2009ix,Adhikari:2016bei}.
They could also be useful in understanding the formation of supermassive 
stars~\cite{VIOLLIER199379,Bilic:2001iv}.
On the other hand, superlight ($\Delta m^2_{41} \sim 10^{-5} \text{ eV}^2$) sterile neutrinos~\cite{deHolanda:2003tx,deHolanda:2010am,Dev:2012bd,Liao:2014ola,Divari:2016jos} may be the explanation for
the lack of upturn in the spectrum of solar neutrino oscillation
probability for energies below $\sim 8$ MeV~\cite{Abe:2016nxk,Aharmim:2011vm,Agostini:2017ixy}.
Recently, it has been pointed out~\cite{deGouvea:2022kma} that a sterile neutrino with $\Delta m^2_{4\ell} \sim 10^{-2} \text{ eV}^2$ (where $\ell$ is the lightest neutrino) can help resolve the tension between the T2K and NOvA data~\cite{Kelly:2020fkv,Esteban:2020cvm,deSalas:2020pgw,Capozzi:2021fjo}.
The question of whether sterile neutrinos exist, and if they do, what their mass and mixing parameters are, is still quite open.

We restrict our attention to the scenario with one sterile neutrino species.
Neutrino oscillation experiments have constrained the mixing angles in the sterile sector ($\theta_{14}, \theta_{24}, \theta_{34}$) over a wide $|\Delta m^2_{41}|$ range~\cite{Acero:2022wqg}. 
However, the identification of the sign of
$\Delta m^2_{41}$ itself has not yet been explored in detail.
Data from cosmology restrict the total amount
of hot dark matter in the Universe and hence constrains the sum of masses of
all neutrinos to $\sum m_i \lesssim O(0.1)$ eV, therefore the sign of $\Delta m^2_{41}$ cannot be negative for 
$|\Delta m^2_{41}| \gtrsim 0.1$ eV$^2$~\cite{Wong:2011ip,FrancoAbellan:2021hdb,Workman:2022ynf}.  However, no such 
constraint has been obtained for smaller $|\Delta m^2_{41}|$ values.
If we were to detect the presence of a sterile neutrino in this mass range, the question of sterile mass ordering --- ``normal'' (Ns) for $\Delta m^2_{41} > 0$ or ``inverted'' (Is) for $\Delta m^2_{41} < 0$ --- would still need to be settled.
\begin{table}[t]
	\centering
	\begin{tabular}{|c|c|c|}
		\hline
		Active mass ordering & Sterile mass ordering & Combination\\
		\hline
		\multirow{2}{*}{$ \Delta m^2_{31} >0$ (N)} & $ \Delta m^2_{41} >0$ (Ns) & N-Ns \\\cline{2-3}
		& $ \Delta m^2_{41} <0$  (Is) & N-Is \\ \hline
		\multirow{2}{*}{ $ \Delta m^2_{31} <0$ (I)} & $ \Delta m^2_{41} >0$  (Ns)& I-Ns \\\cline{2-3}
		& $ \Delta m^2_{41} <0$  (Is) & I-Is \\ \hline
	\end{tabular}
	\caption{All four possible combinations of active and sterile mass ordering}
	\label{tab:NINaIs}
\end{table}
Indeed, since the mass ordering in the active sector (defined by the sign of 
$\Delta m^2_{31}$) as well as the mass ordering in the sterile sector 
(defined by the sign of $\Delta m^2_{41}$) are unknown, we get a total of
4 possible mass ordering combinations as shown in Table~\ref{tab:NINaIs}. The question of mass ordering in the active sector is at the forefront of future physics goals of neutrino experiments. As far as the sterile mass ordering is concerned, it has been shown~\cite{Thakore:2018lgn} that the proposed iron calorimeter (ICAL) experiment at the India-based Neutrino Observatory (INO)~\cite{Kumar:2017sdq} will be sensitive to the sign of $|\Delta m^2_{41}|$ if $|\Delta m^2_{41}| \in (10^{-4}, 10^{-2})$ eV$^2$.
However, such an analysis in the context of long-baseline experiments has never been carried out.

The neutrino oscillation probabilities in the presence of sterile neutrinos in vacuum have been obtained in~\cite{Gandhi:2015xza,Choubey:2017ppj}. Matter effects are included in analytic or semi analytic approaches, in the scenarios where $\Delta m^2_{41}$ is large and hence sterile neutrino oscillations are fast~\cite{Dighe:2007uf,Ray:2010tea,Klop:2014ima,Agarwalla:2016xxa,Haba:2018klh,Sharma:2022qeo}, and for super-light sterile neutrinos where $\Delta m^2_{41}\lesssim\Delta m^2_{21}$~\cite{Liao:2014ola,Divari:2016jos}.
For a wider range of $\Delta m^2_{41}$ encompassing heavy as well as light sterile neutrinos, various approaches for calculating neutrino oscillation probabilities have been employed~\cite{Kamo:2002sj,Li:2018ezt,Parke:2019jyu,Yue:2019qat,Reyimuaji:2019wbn,Fong:2022oim}. However, to explore the complex dependence of sterile oscillations on neutrino mixing parameters, in the presence of matter, one needs to calculate these probabilities with explicit analytic dependence on $\Delta m^2_{41}$, the matter potential and neutrino mixing parameters.

In this paper, we calculate  the conversion probability $P_{\mu e}$ that is valid for all values of $\Delta m^2_{41}$, and has explicit dependence on matter effects.
When calculated as an expansion in the small parameters, the dependence on sterile mixing angles $\theta_{24}$ and $\theta_{34}$ is found to be separable~\cite{Haba:2018klh}. Moreover, the $\theta_{34}$ dependence appears only due to neutral-current forward scattering of neutrinos in matter~\cite{Gandhi:2015xza,Choubey:2017ppj}.
Further, expressing the probability as a summation of terms of the $\sin (x)/x$ form allows the identification of regions in the sterile neutrino parameter space where the combined effect of sterile mixing and matter effect is significant.
This also enables us to explain the features of sterile contribution to $P_{\mu e}$, such as the positions and heights of dips and peaks of $P_{\mu e}$, at a long-baseline neutrino experiment.
 
The analytic expressions calculated in this paper facilitate explorations of many aspects of sterile neutrino oscillations in matter for any possible value of $\Delta m^2_{41}$. In this article, we focus on identifying the sterile mass ordering at a long-baseline neutrino experiment, taking the Deep Underground Neutrino Experiment (DUNE)~\cite{DUNE:2015lol,DUNE:2020ypp,DUNE:2020jqi} as an example. Our analytic expressions indicate that DUNE would be highly sensitive to the mass ordering in the sterile sector in the range of  $|\Delta m^2_{41}| \in (10^{-4}, 10^{-2})$ eV$^2$, where matter effects will play a significant role. We point out key features of sterile neutrino contributions and determine the sensitivity of DUNE to the sign of $\Delta m^2_{41}$. 
We also study how the current uncertainties in the values of other oscillation parameters would affect this sensitivity. We carry out the analyses for all the four mass ordering combinations in table~\ref{tab:NINaIs}, for both neutrinos and antineutrinos.

In section~\ref{sec:analytical}, we present approximate expressions for the
neutrino oscillation probability $P_{\mu e}$ in constant density matter in the presence of a sterile neutrino. 
We analytically explore the sterile mass ordering effects and point out the parameter ranges where these effects will be significant.
In section~\ref{sec:results}, we calculate the sensitivity of DUNE to sterile mass ordering. We also explore the dependence of this sensitivity on $|\Delta m^2_{41}|$ and all four mass ordering combinations.
In section~\ref{sec:concl} we conclude with a discussion on further broader usage of the formalism developed in this paper.

\section{Analytic Approximation for the Conversion Probability $P_{\mu e}$}
\label{sec:analytical}
The upcoming long-baseline neutrino experiment DUNE is primarily sensitive to the conversion channel $\nu_\mu \to \nu_e$.
In this section, we calculate the analytic form for the probability $P_{\mu e} \equiv P(\nu_\mu \to \nu_e)$ in constant density matter, explicitly including the effects of a sterile neutrino of arbitrary mass. Let us first define the Hamiltonian for the $3+1$ neutrino system, in the flavor basis:
\begin{equation}
	\mathcal{H}_{3+1}= \frac{1}{2 E_\nu} \; U \left(
	\begin{array}{cccc}
		0 \quad& 0 & 0 & 0 \\
		0 \quad& \Delta m^2_{21} & 0 & 0 \\
		0 \quad& 0 & \Delta m^2_{31} & 0 \\
		0 \quad& 0 & 0 & \Delta m^2_{41} \\
	\end{array}
	\right) U^\dagger + \left(
	\begin{array}{cccc}
		V_e+V_n & 0 & \quad 0 & \quad 0 \\
		0 & V_n & \quad 0 & \quad 0 \\
		0 & 0 & \quad V_n & \quad 0 \\
		0 & 0 & \quad 0 & \quad 0 \\
	\end{array}
	\right)\;.
\end{equation}
In the above equation,  $V_e \equiv \sqrt{2} G_F N_e$ and $V_n \equiv - G_F N_n / \sqrt{2}$ are the effective charged-current and neutral-current potentials, respectively, experienced by neutrinos due to matter effects. Here, $G_F$ is the Fermi constant and $N_e \; (N_n)$ is electron (neutron) density. The unitary rotation matrix $U$ is parametrized as $U= U_{34} \; U_{24} \; U_{14} \; U_{23} \; U_{13} \; U_{12}\,,$
where each $U_{ij}$ matrix is the unitary rotation matrix in $ij-$plane. The matrix $U$ is expressed in terms 6 independent rotation angles ($\theta_{12}, \theta_{13}, \theta_{23}, \theta_{14}, \theta_{24}, \theta_{34}$) and 3 independent phases ($\delta_{13},\delta_{24},\delta_{34}$).

We define a few dimensionless quantities that will be used frequently in the analysis:
\begin{align}
	\alpha\equiv\frac{\Delta m^2_{21}}{\Delta m^2_{31}},\quad R\equiv \frac{\Delta m^2_{41}}{\Delta m^2_{31}}, \quad  A_e \equiv \frac{2E_\nu V_{e}}{\Delta m^2_{31}},\quad A_n \equiv \frac{2E_\nu V_{n}}{\Delta m^2_{31}}, \quad  \Delta \equiv \frac{\Delta m^2_{31}L}{4 E_\nu}\;.
\end{align}
Our analysis is motivated by the observation that $\theta_{13}$ and $\alpha$ are small quantities and the  active-sterile mixing angles $\theta_{14}$, $\theta_{24}$, and $\theta_{34}$ are also expected to be small. Our approach will consist of perturbative expansions in these small quantities. We define an accounting parameter $\lambda\equiv0.2$ and use
\begin{equation}
	\alpha = 0.03 \sim O(\lambda^2),\qquad s_{13} \simeq 0.14 \sim O(\lambda), \qquad s_{14},\; s_{24},\; s_{34} \sim O(\lambda)\;, \label{eq:orderlambda}
\end{equation}
where $s_{ij} \equiv \sin (\theta_{ij})$.

In order to calculate the probability, we employ the Cayley-Hamilton theorem~\cite{Lindner:2001fx}, which states that any function $g(\mathbb{X})$ of a matrix $\mathbb{X}$ may be expressed as
	\begin{equation}
		g(\mathbb{X})=\sum_{i=1}^k X_i\; g\left(\Lambda _i\right)\; ,\quad \text{ with }\quad X_i \equiv \prod_{j=1,\; j\neq i}^{k} \frac{1}{\Lambda_i-\Lambda_j}(\mathbb{X}-\Lambda_j \mathbb{I})\;.
	\end{equation}
	Here, $\Lambda_i$'s are the distinct eigenvalues of the matrix $\mathbb{X}$. We identify $\mathbb{X}\equiv -i \mathcal{H}_{3+1} L$ so that the probability amplitude matrix in the flavor basis,
	\begin{equation}
		\mathcal{A}_f \equiv \exp (-i \mathcal{H}_{3+1} L) =\exp (\mathbb{X})\;,
	\end{equation}
	can be calculated. This gives the amplitude for oscillation from $\nu_\alpha$ to $\nu_\beta$ as $ A(\nu_\alpha \to \nu_\beta) = [\mathcal{A}_f]_{\beta\alpha} $. The probability is obtained from the amplitude as $P_{\alpha\beta} = |A(\nu_\alpha \to \nu_\beta)|^2$. We first calculate the eigenvalues of $\mathbb{X}$ in the presence of sterile neutrinos, with exact dependence on $\Delta m^2_{41}$ and matter effect, as a perturbative expansion in the small parameters listed in eq.~(\ref{eq:orderlambda}). Using the eigenvalues, we calculate the amplitude $A(\nu_\mu \to \nu_e)$ and the conversion probability $P_{\mu e}$. In the next section, we shall present an explicit expression for the probability calculated up to $O(\lambda^3)$.

\subsection{Decoupling of $\theta_{24}$ and $\theta_{34}$ -dependent terms in matter}
The analytic expression for the conversion probability $P_{\mu e}$, correct up to $O(\lambda^3)$, is
\begin{align}
	P_{\mu e}=&\; 4 \, s_{13}^2 \, s_{23}^2 \, \frac{\sin ^2  \left[ \left(A_e-1\right) \Delta \right]}{\left(A_e-1\right)^2} \nonumber \\
	&+2 \, \alpha \, s_{13} \,  \sin 2 \theta _{12} \sin 2 \theta _{23} \cos \left(\delta _{13}+\Delta \right) \frac{\sin  \left[\left(A_e-1\right) \Delta\right]}{A_e-1} \frac{\sin \left[A_e \Delta\right]}{A_e} \nonumber\\
	&+4 \, s_{13} \,  s_{14} \, s_{24} \, s_{23}  \frac{\sin \left[\left(A_e-1\right) \Delta\right]}{A_e-1}\nonumber\\
	&\quad\times R \Bigg[ A_n s_{23}^2 \frac{\sin \left[\left(A_e-1\right)\Delta+\delta _{24}^\prime \right]}{\left(A_e-1\right) \left(A_n+1-R\right)} +A_n c_{23}^2 \frac{\sin \left[  \left(A_e+1\right)\Delta+\delta_{24}^\prime\right]}{A_e \left(A_n-R\right)} \Bigg.\nonumber\\
	&\quad\qquad- \left(R-\frac{A_n s_{23}^2 }{A_n+1-R}\right)\frac{\sin \left[  \left(A_e+2 A_n-2 R+1\right)\Delta+\delta _{24}^\prime \right]}{\left(A_n-R\right) \left(A_e+A_n-R\right)} \nonumber\\
	&\quad\qquad\Bigg. - \left(c_{23}^2 \frac{A_n }{A_e}-\left[A_e+A_n-1\right]\right) \frac{\sin \left[\left(A_e-1\right)\Delta-\delta _{24}^\prime \right]}{\left(A_e-1\right) \left(A_e+A_n-R\right)}\Bigg]\nonumber\\
	&+4 s_{13} \, s_{14}  \, s_{34} \, s_{23}^2 \,  c_{23} \, A_n \frac{\sin\left[ \left(A_e-1\right) \Delta\right] }{A_e-1}\nonumber\\
	&\quad\times R \Bigg[ \frac{\sin \left[\Delta  \left(A_e-1\right)+\delta_{34}^\prime \right]}{\left(A_e-1\right) \left(A_n+1-R\right)} + \frac{\sin \left[\left(A_e+2 A_n-2 R+1\right)\Delta+\delta _{34}^\prime\right]}{\left(A_n+1-R\right) \left(A_n-R\right) \left(A_e+A_n-R\right)} \Bigg.\nonumber\\
	&\quad\qquad \Bigg.-\frac{\sin \left[\left(A_e+1\right)\Delta+\delta _{34}^\prime \right]}{A_e \left(A_n-R\right)}+\frac{\sin \left[\left(A_e-1\right)\Delta-\delta _{34}^\prime \right]}{A_e \left(A_e-1\right)  \left(A_e+A_n-R\right)}\Bigg] +O(\lambda^4)\;.
	\label{eq:Pmue}
\end{align}
Here, $\delta_{24}^\prime $ and $\delta_{34}^\prime $ are defined as $	\delta_{24}^\prime \equiv \delta_{24}+\delta_{13}$ and $ \delta_{34}^\prime \equiv \delta_{34}+\delta_{13}$, since both $\delta_{24}$ and $\delta_{34}$ appear only in this combination. Note that the above expression includes the exact dependence on $R\equiv \Delta m^2_{41}/ \Delta m^2_{31}$ (i.e. on $ \Delta m^2_{41}$) as well as on the constant density matter potentials $A_e$ and $A_n$.

The first two terms in eq.~(\ref{eq:Pmue}) are simply the three-neutrino ($3\nu$) contributions to $P_{\mu e}$~\cite{Akhmedov:2004ny}, whereas the last two terms are the contributions due to sterile neutrinos. Note that the $\theta_{24}$ dependence appears only in the third term, and the $\theta_{34}$ dependence appears only in the fourth term. Thus, these two contributions are decoupled as long as the assumption of the smallness of the sterile mixing angles is valid. While the $\theta_{24}$ contribution is present even in the vacuum limit, the $\theta_{34}$ dependent term is non-zero when $A_n \neq 0$, i.e. only in the presence of matter effects. The $\theta_{34}$ contribution may be observed to be suppressed by a factor of $s_{23}\, c_{23} \simeq 0.5$ as compared to the $\theta_{24}$ contribution.

We also observe that the sterile neutrino contributions are regulated by 
\begin{equation}
	\sin \left[\left(A_e-1\right)\Delta\right]/(A_e-1)\;.
\end{equation}
This dependence also appears in the first two terms in eq.~(\ref{eq:Pmue}) that represent the contributions from the active neutrino sector. Thus a significant contribution from the sterile oscillation is expected to be present near $|\left(A_e-1\right)\Delta|=\pi/2$, i.e. near the first oscillation peak while approaching from higher energies. Further dependence of the sterile oscillation peaks on $\Delta m^2_{41}$ and matter effects will be discussed throughout this paper.

The probability for $P_{\bar{\mu}\bar{e}} \equiv P(\bar{\nu}_\mu \to \bar{\nu}_e)$ is obtained by the replacements
\begin{equation}
	A_e \to -A_e\;,\qquad A_n \to -A_n\;,\qquad \delta_{ij} \to -\delta_{ij}\;.
\end{equation}
Note that the analytic expression in eq.~(\ref{eq:Pmue}) is a perturbative expansion in $\alpha$, therefore the expression is only valid for $\alpha \Delta  \lesssim 1$, i.e. when the distance travelled by the neutrinos is much less than the wavelengths of oscillation due to $\Delta m^2_{21}$. For long-baseline and atmospheric neutrino experiments, this is a valid approximation.

\subsection{$P_{\mu e}$ in the $\sin(x)/x $ form, with $A_n=-A_e/2$}
It may be observed that the terms in eq.~(\ref{eq:Pmue}) consist of many quantities with the functional form $\sin (x)/x$. This function reaches a maximum in the limit $x \to 0$. Therefore it is expected that the contribution of such terms will be significant when the corresponding denominator vanishes, without giving rise to any unphysical singularities.  Re-structuring the probability expression in eq.~(\ref{eq:Pmue}) as a summation of $\sin (x)/x$ terms will allow us to identify the regions where certain contributions will be dominant.

Further, for the Earth's crust, we can take $A_n \approx -A_e/2$. This is a very good approximation, since the neutral current and the charged current potentials are related via
\begin{equation}
	A_n =- \frac{A_e}{2} \frac{N_n}{N_e}\;,
\end{equation}
and the number of neutrons and electrons are approximately equal for lighter elements. The restructured expression is
\begin{align}
	P_{\mu e}=&\; 4 \, s_{13}^2 \, s_{23}^2 \frac{\sin ^2  \left[ \left(A_e-1\right) \Delta\right] }{\left(A_e-1\right)^2} \nonumber \\
	&+2 \, \alpha \,  s_{13} \, \sin 2 \theta _{12} \sin 2 \theta _{23} \cos \left(\delta _{13}+\Delta \right) \frac{\sin  \left[\left(A_e-1\right) \Delta\right]}{A_e-1} \frac{\sin \left[A_e \Delta\right]}{A_e}\nonumber\\
	&+4 \, s_{13} \,  s_{14}  \, s_{24} \, s_{23} \, \frac{\sin \left[\left(A_e-1\right)\Delta\right]}{A_e-1} \Big[\sin( \delta_{24}^\prime) P_{24}^{s} + \cos (\delta_{24}^\prime )P_{24}^{c}\Big]\nonumber\\
	&+4 \, s_{13} \,  s_{14} \, s_{34} \,  s_{23}^2 \, c_{23} \, \frac{\sin \left[\left(A_e-1\right)\Delta\right]}{A_e-1} \Big[\sin (\delta_{34}^\prime) P_{34}^{s} + \cos (\delta_{34}^\prime) P_{34}^{c}\Big]+O(\lambda^4)\;, \label{eq:Pmue2}
\end{align}
where the quantities $P_{24}^s$, $P_{24}^c$, $P_{34}^s$, $P_{34}^c$ can be written using expressions of the $\sin(x)/x$ form as follows. The coefficients of $\sin \delta_{24}^\prime$ and $\cos \delta_{24}^\prime$ terms are, respectively,
\begin{align}
	P_{24}^s=&\;R\Big[\tfrac{1}{2} A_e c_{23}^2 +(R-1) \left(s_{23}^2+1\right)\Big]\frac{\sin   \left[\left(R-1+\frac{A_e}{2}\right)\Delta\right]}{R-1+\frac{A_e}{2}}\frac{\sin\left[ \left(R-\frac{A_e}{2}\right) \Delta\right] }{R-\frac{A_e}{2}}\nonumber\\
	&+c_{23}^2 R \sin \left[\left(R-1-\tfrac{A_e}{2}\right)\Delta\right] \frac{\sin \left[\left(R+\frac{A_e}{2}\right) \Delta\right] }{R+\frac{A_e}{2}}\; , \label{eq:Ps24} \\
	P_{24}^c=&\; \frac{R}{R-\frac{1}{2}}\left( \Big[R-\tfrac{1}{2}s_{23}^2-\tfrac{1}{2}\Big] \cos \left[  \left(R-1+\tfrac{A_e}{2}\right)\Delta\right]\frac{\sin \left[  \left(R-\frac{A_e}{2}\right)\Delta\right]}{R-\frac{A_e}{2}}\right.\nonumber\\
	&\left.+s_{23}^2 \frac{\sin \left[\left(A_e-1\right)\Delta\right]}{A_e-1}+ s_{23}^2 (R-1) \cos \left[  \left(R-\tfrac{A_e}{2}\right)\Delta\right]\frac{\sin \left[  \left(R-1+\frac{A_e}{2}\right)\Delta\right]}{R-1+\frac{A_e}{2}} \right)\nonumber\\
	&+c_{23}^2 R \cos \left[\left(R-1-\tfrac{A_e}{2}\right)\Delta \right]\frac{\sin \left[ \left(R+\frac{A_e}{2}\right)\Delta \right]}{R+\frac{A_e}{2}}\; .
	\end{align}
Similarly, the coefficients of $\sin \delta_{34}^\prime$ and $\cos \delta_{34}^\prime$ terms are, respectively,
	\begin{align}
	P_{34}^s=&\; R \Big( R-1-\tfrac{A_e}{2} \Big)  \frac{\sin \left[ \left(R-1+\frac{A_e}{2}\right)\Delta\right]}{R-1+\frac{A_e}{2}} \frac{\sin \left[  \left(R-\frac{A_e}{2}\right)\Delta\right]}{R-\frac{A_e}{2}}\nonumber\\
	&-R \sin \left[\left(R-1-\tfrac{A_e}{2}\right)\Delta\right]\frac{\sin \left[  \left(R+\frac{A_e}{2}\right)\Delta\right]}{R+\frac{A_e}{2}}\; ,\\
	P_{34}^c=&\; \frac{R}{R-\frac{1}{2}}\left( \frac{\sin \left[  \left(A_e-1\right)\Delta\right]}{A_e-1}-\tfrac{1}{2} \cos \left[  \left(R-1+\tfrac{A_e}{2}\right)\Delta\right]\frac{\sin \left[  \left(R-\frac{A_e}{2}\right)\Delta\right]}{R-\frac{A_e}{2}} \right.\nonumber\\
	&\left.+(R-1) \cos \left[  \left(R-\tfrac{A_e}{2}\right)\Delta\right]\frac{\sin \left[  \left(R-1+\frac{A_e}{2}\right)\Delta\right]}{R-1+\frac{A_e}{2}}\right) \nonumber \\
	&-R \cos \left[\left(R-1- \tfrac{A_e}{2}\right)\Delta\right]\frac{\sin \left[ \left(R+\frac{A_e}{2}\right)\Delta\right]}{R+\frac{A_e}{2}}\; .
	\label{eq:Pc34}
\end{align}
From the above expressions, we immediately observe that for $R=1-A_e/2$ and $R=\pm A_e/2$, the sterile neutrino contribution to $P_{\mu e}$ will be enhanced due to resonance in matter.

In the vacuum limit the $\theta_{34}$ dependence vanishes, i.e. $P^s_{34}|^\text{vac} = 0$ and $P^c_{34}|^\text{vac}=0$. On the other hand,
the $P_{24}^s$ and the $P_{24}^c$ terms can be expressed as
\begin{align}
	P^s_{24}|^\text{vac} \simeq &\,  2 \sin[ (R-1) \Delta] \sin [R \Delta]\;,\nonumber\\
	P^c_{24}|^\text{vac} \simeq &\,  2 \cos[ (R-1) \Delta] \sin [R \Delta]\;.
\end{align}
In vacuum these simple expressions can give the positions of the peaks and dips due to sterile neutrino oscillations. For a given value of $R$ chosen by Nature, matter effects will be significant around $A_e \approx \pm 2R$ and $A_e = 2(1-R)$ and will modify the values of $P^s_{24}$, $P^c_{24}$, $P^s_{34}$, and $P^c_{34}$.

The insights from the eqs.~(\ref{eq:Ps24})-(\ref{eq:Pc34}) in this section determine the positions and amplitudes of peaks and dips in $P_{\mu e}$, and can be used to probe the sensitivity of long baseline experiments to sterile neutrino mass ordering.

\subsection{Effects of sterile mass ordering on $P_{\mu e}$ at DUNE}

\begin{figure}[t]\centering
	\includegraphics[width=0.47\textwidth]{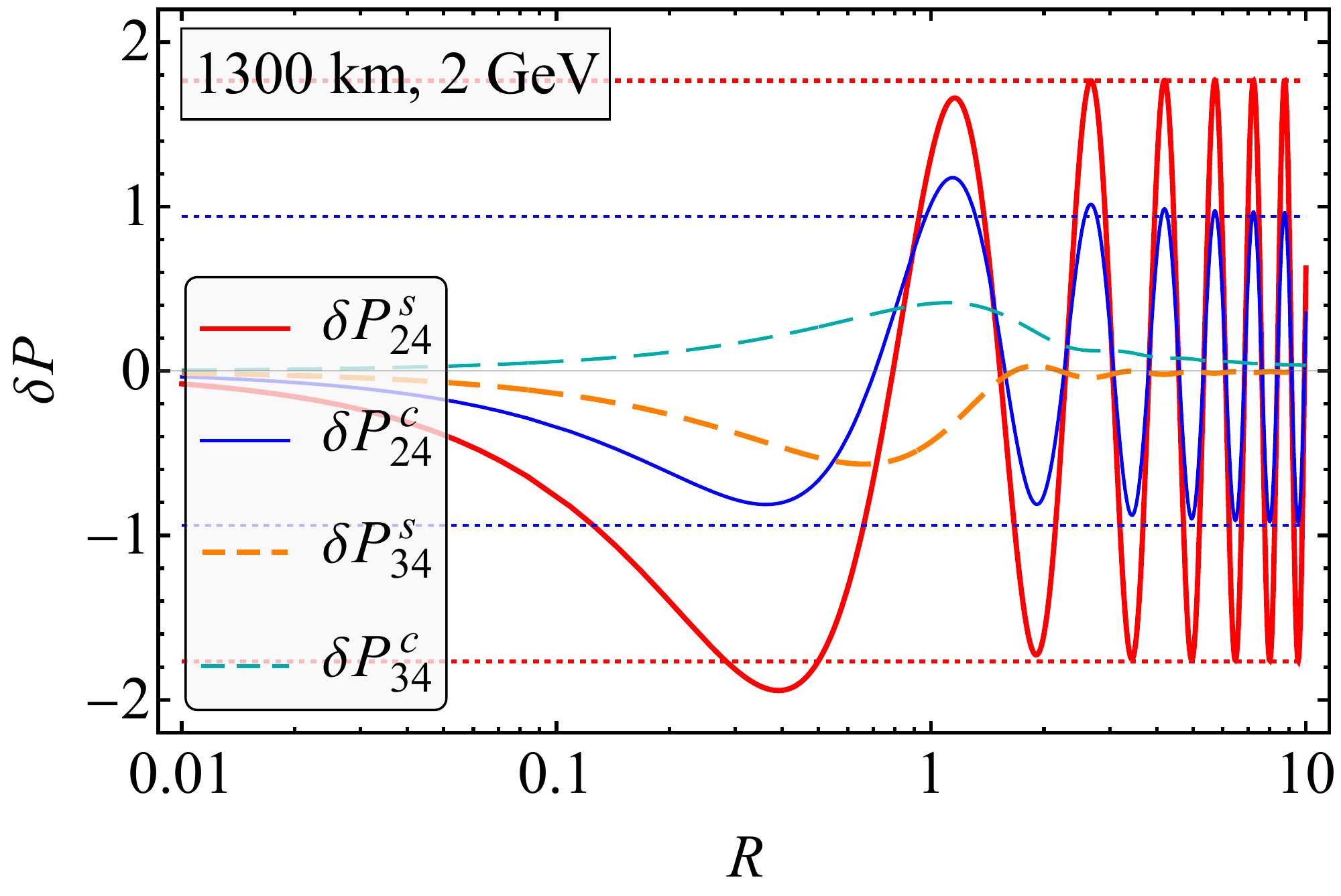} \hspace{5pt}
	\includegraphics[width=0.47\textwidth]{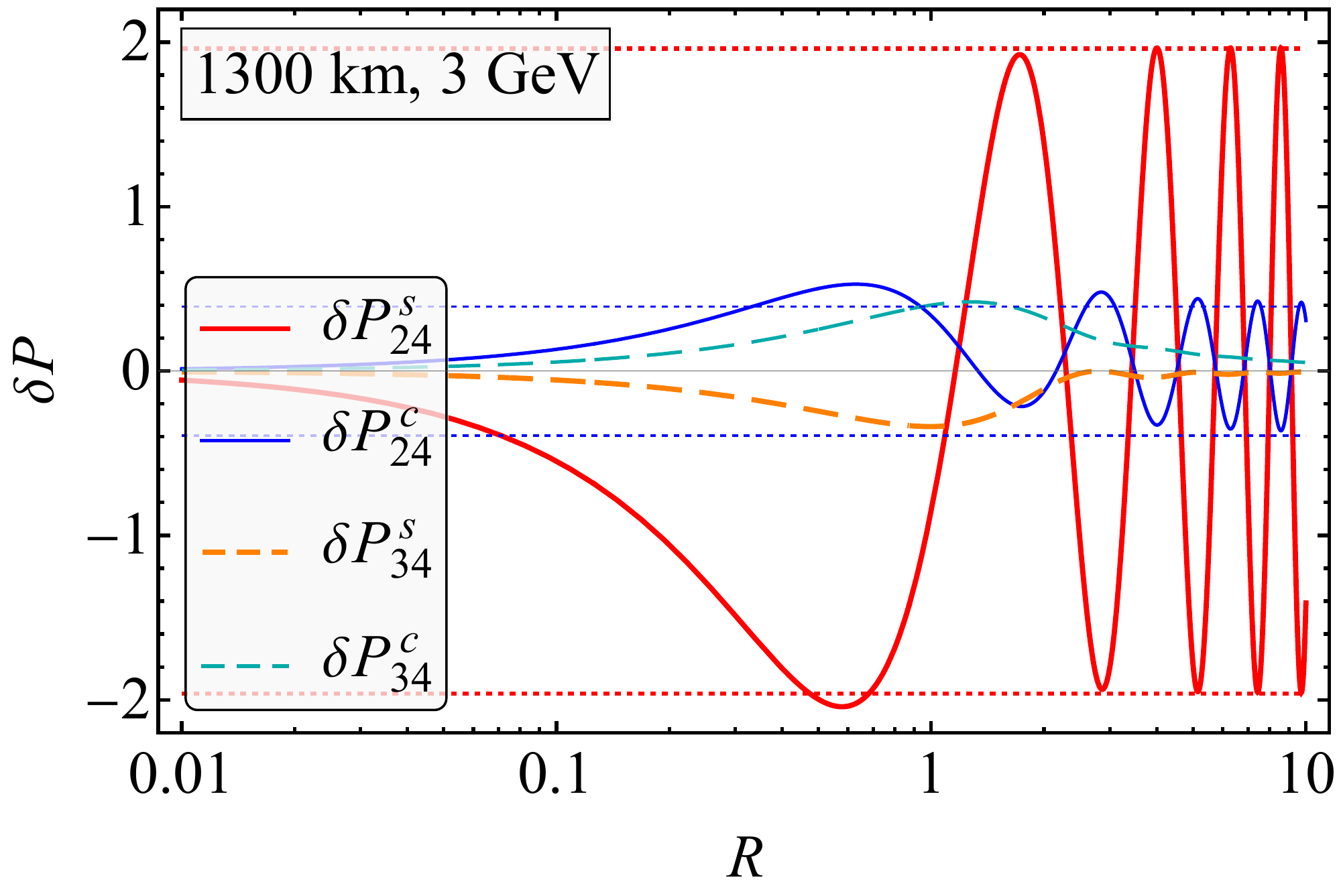}
	\caption{The values of $\delta \! P^s_{24}$, $\delta \!  P^c_{24}$, $\delta \!  P^s_{34}$ and $\delta \!  P^c_{34}$ as functions of $R=\Delta m^2_{41}/\Delta m^2_{31}$, at $E_\nu=2$ GeV [left] and $E_\nu=3$ GeV [Right], for DUNE. The Horizontal dotted line are the bounds on the amplitude of $\delta \! P^s_{24}$ and $\delta \!  P^c_{24}$ in the vacuum limit, obtained from eq.~(\ref{eq:vacdelP}). }
	\label{fig:DuneDeltaP}
\end{figure}

The sensitivity of the oscillation probability $P_{\mu e}$ to the mass ordering in the sterile sector may be examined using the quantity
\begin{equation}
	\delta \! P_{\mu e} = P_{\mu e} (R)- P_{\mu e}(-R)\;.
\end{equation}
This quantity clearly depends on $\delta \! P^s_{24}$, $ \delta \! P^c_{24}$, $\delta \! P^s_{34}$ and $\delta \! P^c_{34}$ where `$\delta$' indicates the difference between the values of these quantities for positive and negative values of $R$. Indeed,
 \begin{align}
	\delta \! P_{\mu e} = \;	&4 \, s_{13} \,  s_{14}  \, s_{24} \, s_{23} \, \frac{\sin \left[\left(A_e-1\right)\Delta\right]}{A_e-1} \Big[\sin( \delta_{24}^\prime) \, \delta \! P_{24}^{s} + \cos (\delta_{24}^\prime ) \, \delta \! P_{24}^{c}\Big]\nonumber\\
	&+4 \, s_{13} \,  s_{14} \, s_{34} \,  s_{23}^2 \, c_{23} \, \frac{\sin \left[\left(A_e-1\right)\Delta\right]}{A_e-1} \Big[\sin (\delta_{34}^\prime) \, \delta \!  P_{34}^{s} + \cos (\delta_{34}^\prime) \, \delta \! P_{34}^{c}\Big]\;.
	\label{eq:sterilecontr}
\end{align}
To understand the contributions of various $\delta \! P$'s, we plot their values at $E_\nu=2\text{ and }3$ GeV in figure~\ref{fig:DuneDeltaP}. These energies are in the range where the flux of DUNE is near its maximum and where the first oscillation peak is expected to be observed. 
It may be seen that all four quantities have significant non-zero values, depending on the value of $R$.

Note that the quantities $\delta \! P^s_{34}$ and $\delta \! P^c_{34}$, which are expected to vanish in the vacuum limit, are non-zero due to the inclusion of matter effects. 
While $\delta \! P^s_{24}$ and $\delta \! P^c_{24}$ are non-zero in vacuum:
\begin{equation}
	 	\delta \! P^s_{24}(\text{vac})= -2  \sin (2 R \Delta) \sin (\Delta )\;, \qquad \delta \! P^c_{24}(\text{vac} )= +2  \sin (2 R \Delta) \cos (\Delta )\;, \label{eq:vacdelP}
\end{equation}
their detailed behavior is affected by matter effects. For example, the first oscillation peak would be at $\Delta =\pi/2$ in vacuum, where $ \delta \! P^c_{24}(\text{vac} )$ would vanish. However in matter, the first oscillation peak can be approximated to be at $(1-A_e)\Delta =\pi/2$.
Since
\begin{equation}
	A_e \approx 2.95 \times 10^{-2} \Big(\frac{E_\nu}{1\text{ GeV}}\Big) \Big(\frac{\rho}{1 \text{ g/cc}} \Big)\;,
\end{equation}
we have $A_e \simeq 0.2$, for $E_\nu \sim 2.5$ GeV at DUNE.
As a result, both $\delta\! P^s_{24}$ and $\delta\! P^c_{24}$  will be non-zero at the first oscillation peak.
From eq.~(\ref{eq:sterilecontr}), this implies that DUNE would be sensitive\footnote{Note that for T2K and NOVA, due to smaller matter effects,  $\delta\! P^c_{24}$ would be small at all oscillation peaks. This leads to a strong dependence of the sterile mass ordering sensitivity to the value of $\delta_{24}^\prime$.} to sterile mass ordering for all possible values of $\delta_{24}^\prime$.

Note that the effects of matter-induced resonance at $A_e =2(1-R)$ and $A_e = \pm 2R$ cannot be accounted for by the simple vacuum limit approximations given above in eq.~(\ref{eq:vacdelP}). The effects of these resonances will be discussed in detail in section~\ref{sec:results}.

For our analysis, we choose the benchmark parameters in the 3$\nu$ sector to be
\begin{align}
	|\Delta m^2_{31}| = 2.5\times 10^{-3}\text{ eV}^2\;,\qquad & \Delta m^2_{21} = 7.5\times 10^{-5}\text{ eV}^2\;, \qquad \delta_{13}=-90^\circ\;, \nonumber\\
	\theta_{12}=33.56^\circ \;,\qquad & \theta_{23}=45^\circ\;,\qquad \theta_{13}=8.46^\circ\;.
	\label{eq:activemixing}
\end{align}
This is consistent with the global fits~\cite{Esteban:2020cvm,nufit}.
We choose the sterile sector parameter values as
\begin{align}
	\theta_{14}=5^\circ\;,\quad \theta_{24}=10^\circ\;, \quad \theta_{34}=0^\circ\;,\quad \delta_{24}=0^\circ\;,\quad \delta_{34}=0^\circ\;. \label{eq:sterilemixing}
\end{align}
With the above choice, only the $P^s_{24}$ contribution due to sterile neutrinos will stay. This simplifies the analytic exploration of the features of $P_{\mu e}$. Later in section~\ref{sec:results}, we shall find that many features in sensitivity to sterile mass ordering as a function of $|\Delta m^2_{41}|$ can be explained by observing the behavior of $P^s_{24}$ in matter.

\begin{figure}[t]\centering
	\hspace{-10mm}
	\begin{subfigure}{.94\textwidth}
		\centering
		\includegraphics[width=0.48\textwidth]{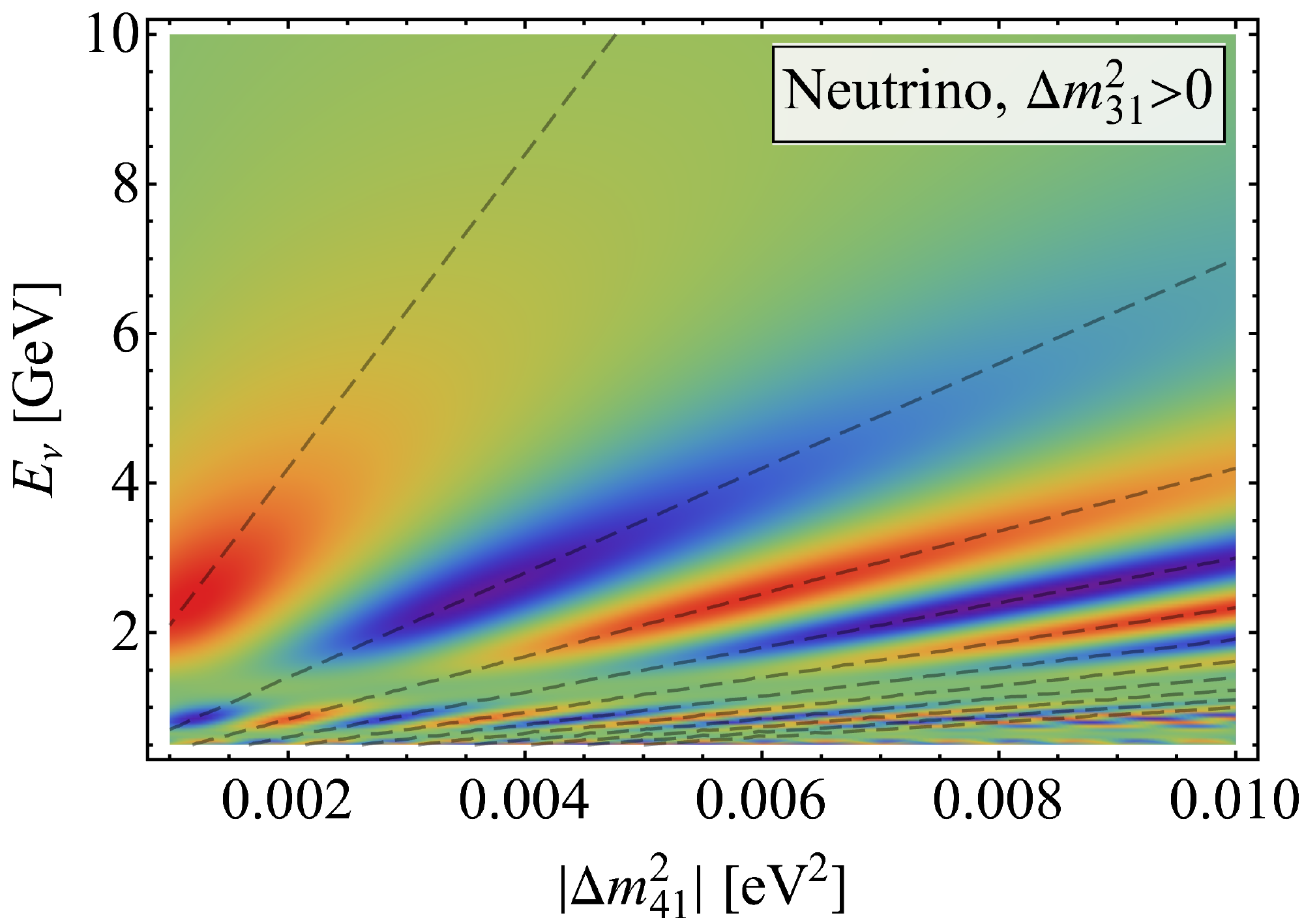}
		\includegraphics[width=0.48\textwidth]{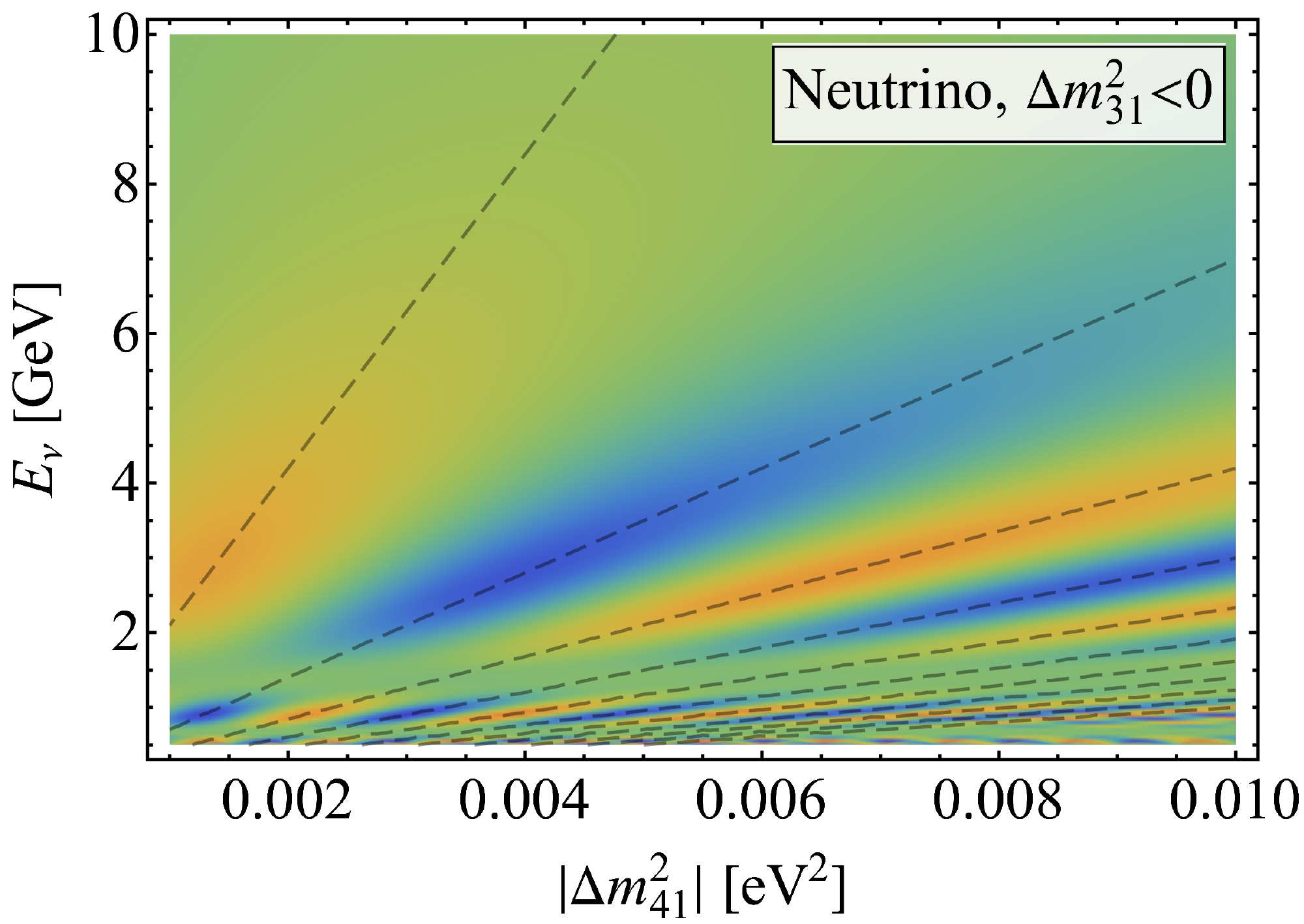}\\
		\vspace{10pt}
		\includegraphics[width=0.48\textwidth]{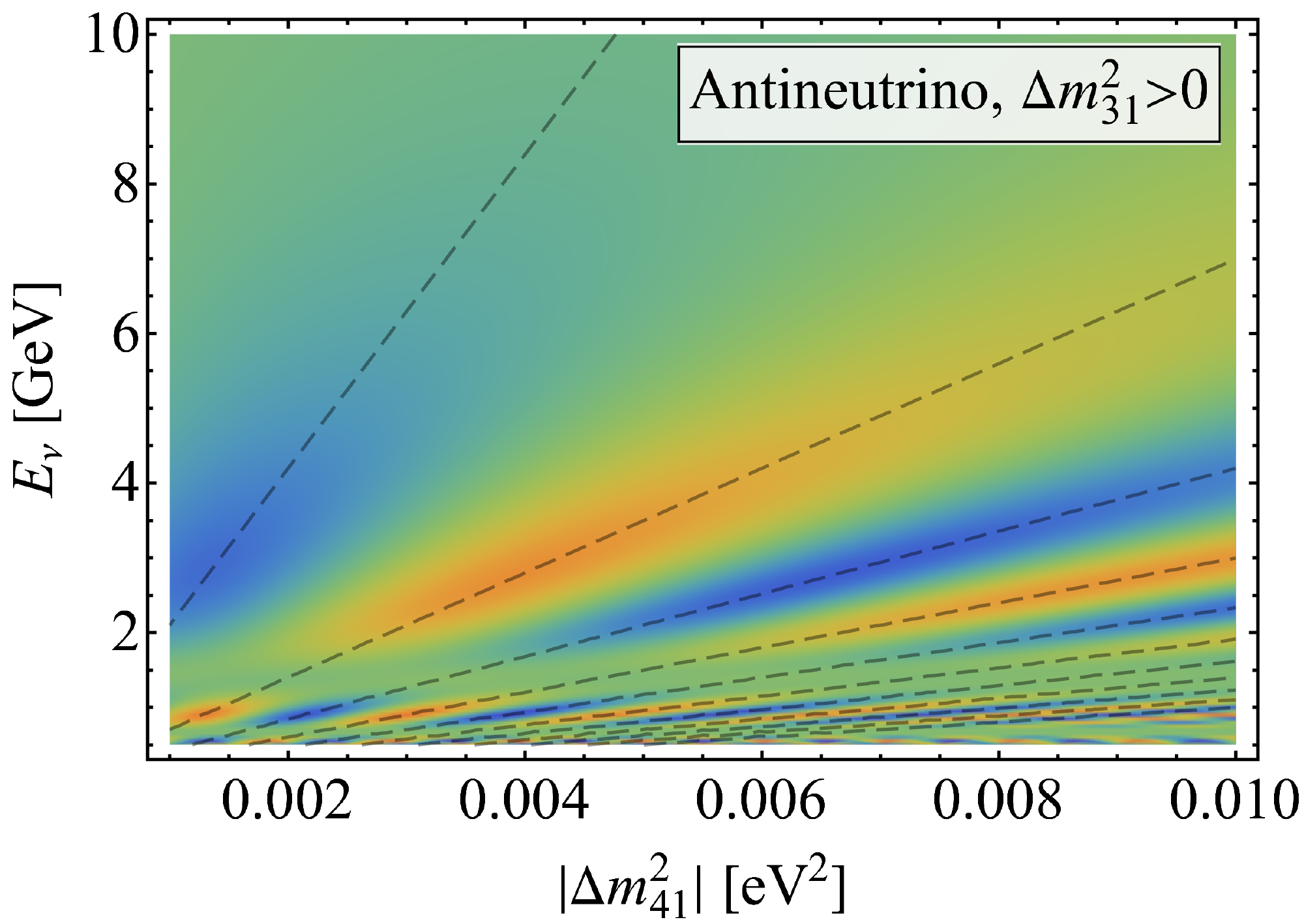}
		\includegraphics[width=0.48\textwidth]{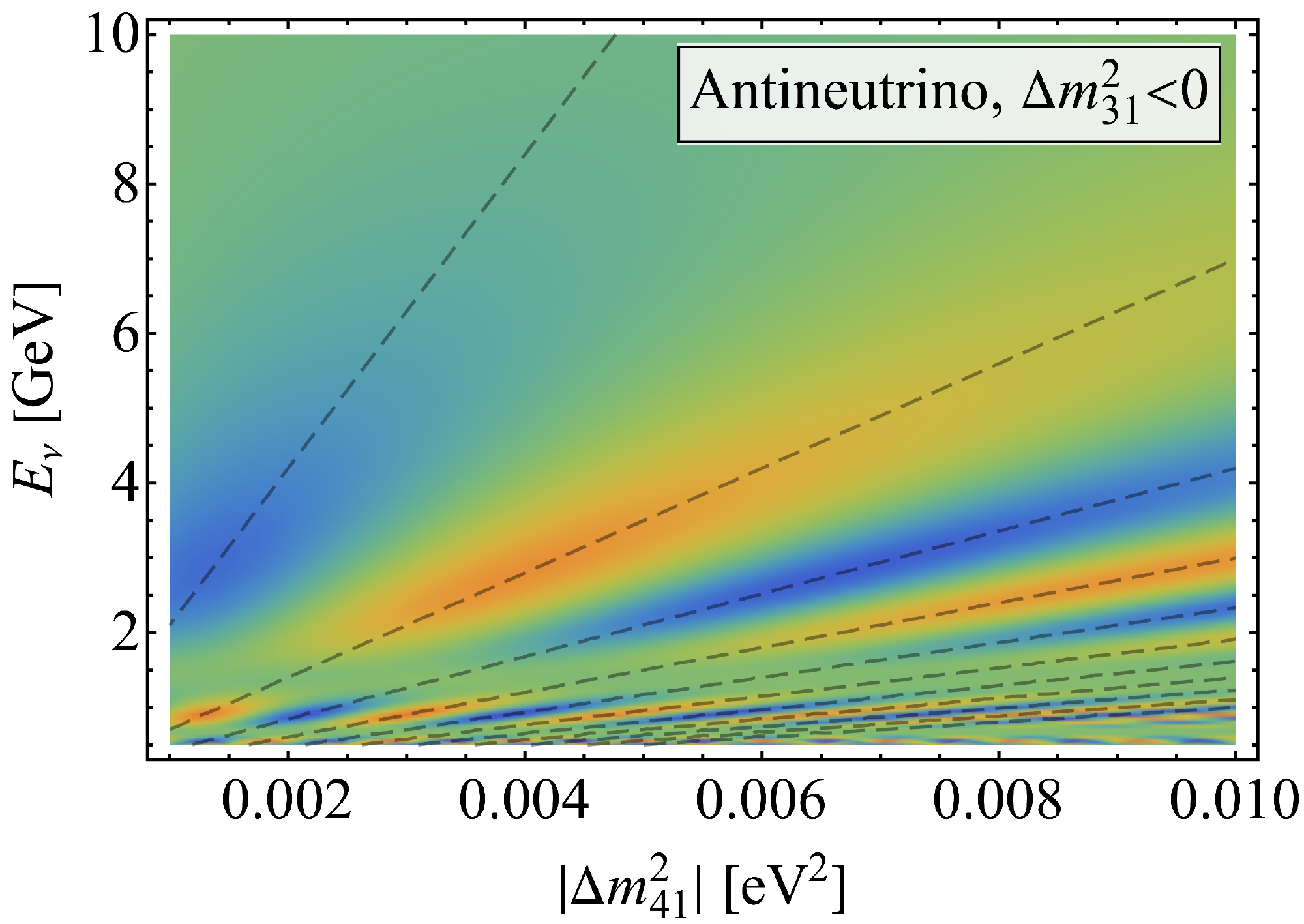}
	\end{subfigure}
	\begin{subfigure}{.0265\textwidth}
		\includegraphics[width=3\textwidth]{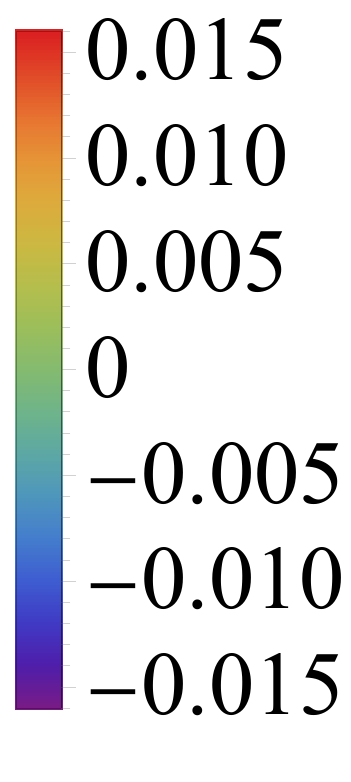}
	\end{subfigure}
	\caption{$\delta \! P_{\mu e}$ [Top panels] and $\delta \! P_{\bar{\mu} \bar{e}}$ [Bottom panels] in the $(\Delta m^2_{41}-E_\nu)$ plane at $L=1300$ km for the mixing parameters given in eqs.~(\ref{eq:activemixing})-(\ref{eq:sterilemixing}). Left and Right panels correspond to Normal and Inverted mass ordering, respectively, in the active sector.}
	\label{fig:deltapmue}
\end{figure}

We plot $\delta \! P_{\mu e}$ and $\delta \! P_{\bar{\mu} \bar{e}}$ in the $(\Delta m^2_{41},E_\nu)$ plane in figure~\ref{fig:deltapmue}, and observe that the sensitivity to sterile mass ordering depends on whether we are observing neutrinos or anti-neutrinos, as well as on the sign of $\Delta m^2_{31}$ (normal or inverted mass ordering of active neutrinos).
The following observations may be made from figure~\ref{fig:deltapmue}
\begin{itemize}
	\item The values of $|\delta \! P_{\mu e}|$ and $|\delta \! P_{\bar\mu \bar e}|$ are observed to be maximum at $E_\nu \sim 2-3$ GeV. This is primarily due to the $\sin[(1-A_e)\Delta]/(1-A_e)$ dependence of the sterile contribution, as obtained in eq.~(\ref{eq:sterilecontr}).
	\item The peaks  and valleys of $\delta \! P_{\mu e}$ and $\delta \!  P_{\bar{\mu }\bar{e}}$ correspond approximately to the $\sin(2R\Delta)$ dependence of the $\delta \! P^s_{24}$ term in eq.~(\ref{eq:vacdelP}). This dependence is represented by the black dashed lines in figure~\ref{fig:deltapmue}.
	\item For higher values of $\Delta m^2_{41}$ (i.e. $R\gg 1$), we observe the expected rapid oscillation at low energies.
	\item The amplitudes of the peaks and dips are maximum for neutrino with $\Delta m^2_{31}>0$. 
	\item The locations of peaks and valleys approximately interchange between $\nu$ and $\bar{\nu}$ plots. This is because the only non-zero contribution to $|\delta \! P_{\mu e}|$ is from $\delta \! P^s_{24}$, which is the coefficient of $\sin (\delta_{24}^\prime)$, and $\delta_{ij}\to -\delta_{ij}$ when $\nu \to \bar\nu$.
\end{itemize}
When a non-zero contribution of $P^c_{24}$ is present, we expect the dependence of $\delta \! P_{\mu e}$ on $\Delta m^2_{41}$ and $E_\nu$ to change, however eqs.~(\ref{eq:Ps24})-(\ref{eq:Pc34}) can explain the dominant characteristics of $\delta \! P_{\mu e}$ in such a scenario.

\subsection{Peaks and dips in $P_{\mu e}$ at DUNE due to sterile neutrino}

\begin{figure}[t]\centering
	\includegraphics[width=0.48\textwidth]{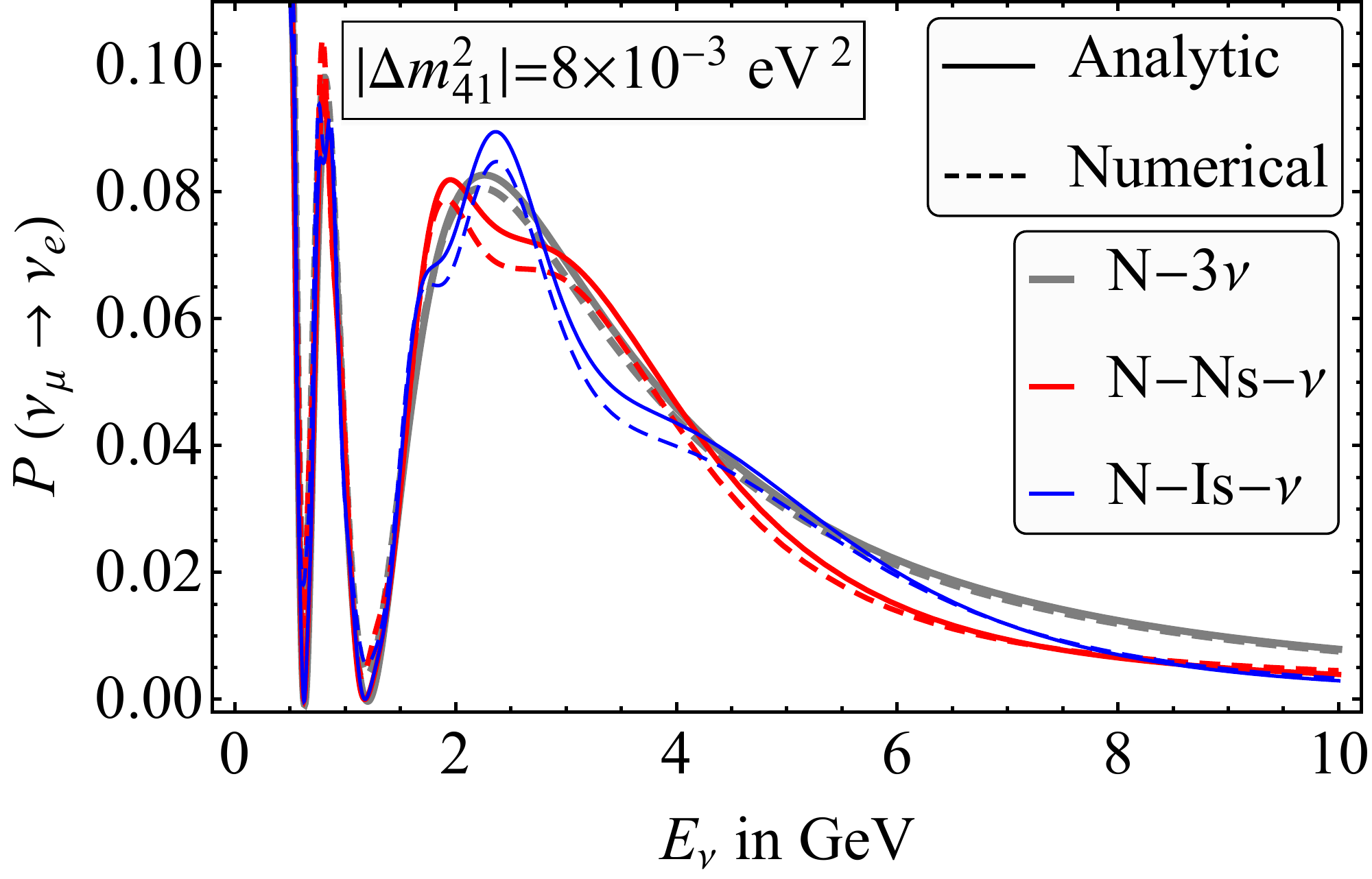}
	\includegraphics[width=0.48\textwidth]{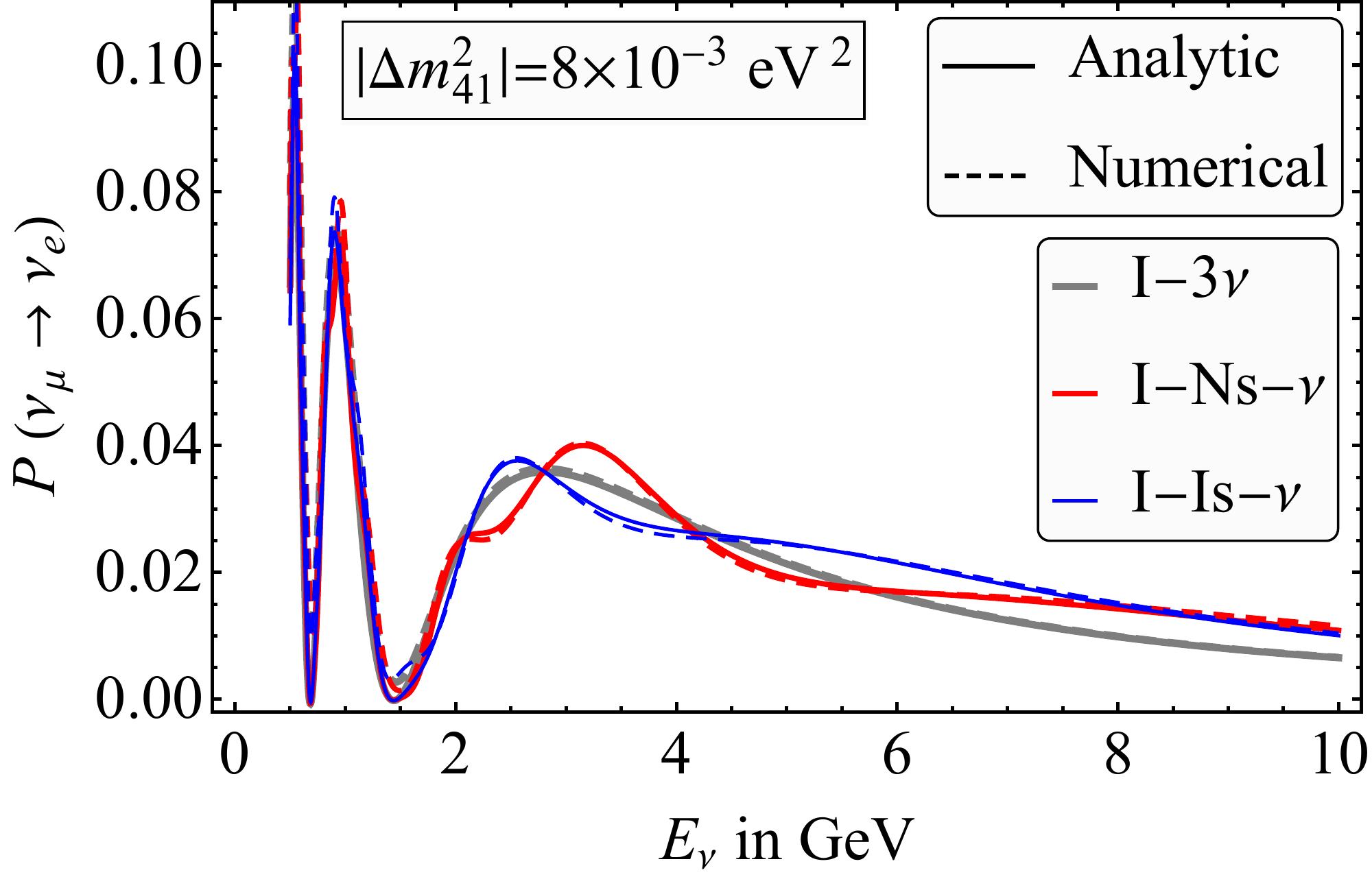}\\
	\vspace{10pt}
	\includegraphics[width=0.48\textwidth]{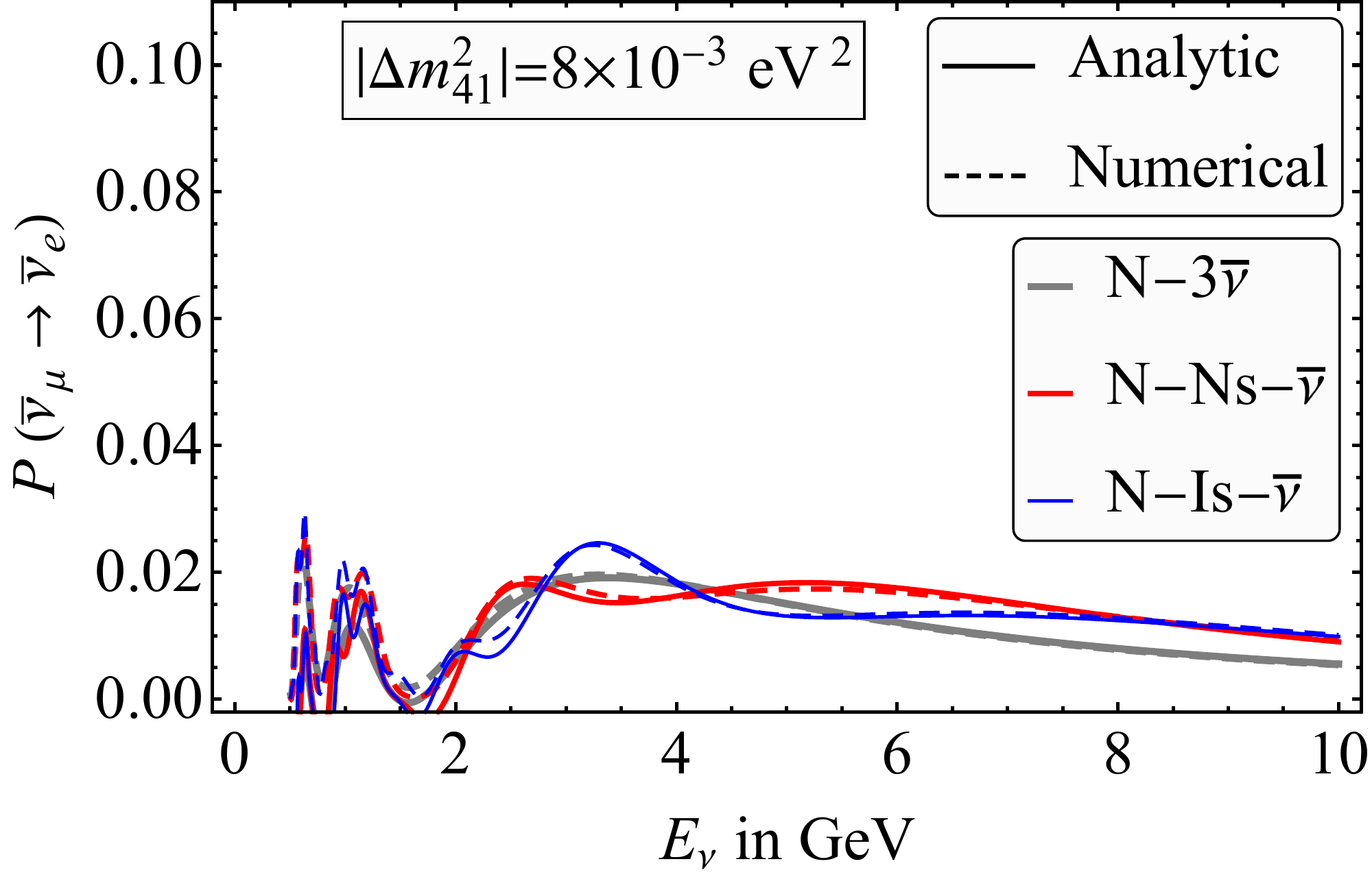}
	\includegraphics[width=0.48\textwidth]{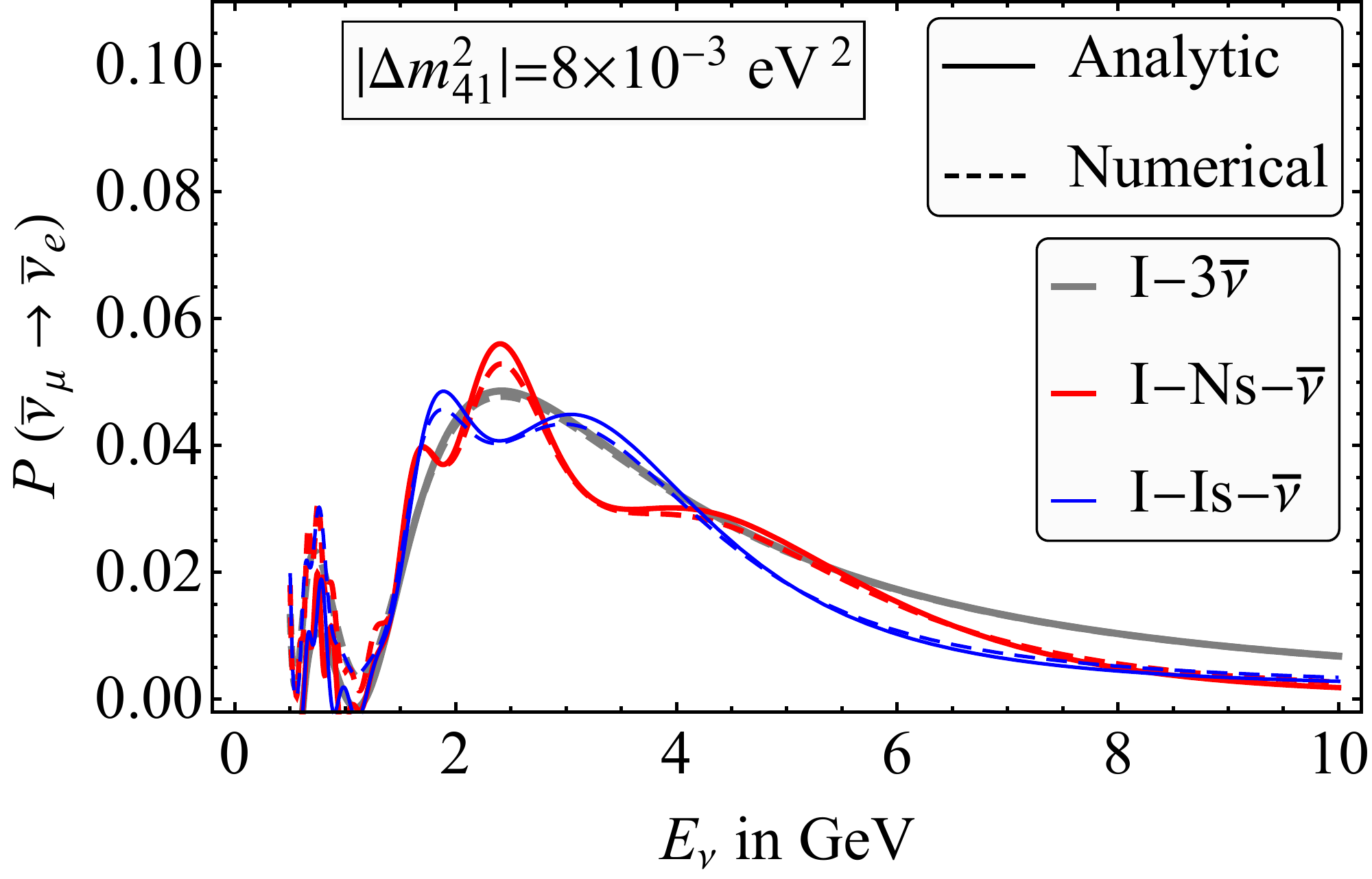}
	\caption{Conversion probabilities $P(\nu_\mu \to \nu_e)$ [Top panels] and $P(\bar{\nu}_\mu \to \bar{\nu}_e)$ [Bottom panels] as functions of energy ($E_\nu$) for $L=1300$ km, for the mixing parameters given in eqs.~(\ref{eq:activemixing})-(\ref{eq:sterilemixing}) and for $| \Delta m^2_{41}| = 8 \times 10^{-3} \text{ eV}^2$. Left and Right panels correspond to Normal and Inverted mass ordering, respectively, in the active sector.}
	\label{fig:comparison}
\end{figure}

The analytic expressions  in Eqs.~(\ref{eq:Pmue2})-(\ref{eq:Pc34}) can explain the features of sterile neutrino contributions to $P_{\mu e}$ quite well, as can be seen in figure~\ref{fig:comparison}. We choose $| \Delta m^2_{41}| = 8 \times 10^{-3} \text{ eV}^2$ ($R=3.2$) for comparison between the numerical and analytic solutions.
Plotting $P(\nu_\mu \to \nu_e)$ and $P(\bar\nu_\mu \to \bar\nu_e)$ for normal and inverted mass ordering (in both the sectors) in figure~\ref{fig:comparison}, we observe that the two sterile mass orderings lead to distinctly different shapes of the conversion probability.
For both the neutrino and antineutrino channels, with normal or inverted mass ordering (in the active or the sterile sector), our analytic approximations follow the exact numerical results with an absolute accuracy of better than $\sim 1\%$. Typically the sterile contribution results in additional
peaks and dips which are more visible near the first oscillation peak of the $3\nu$ sector. 
The positions and the amplitudes of the sterile as well as the $3\nu$ peaks and dips are observed to be reproduced extremely well.
	
The positions and amplitudes of these peaks and dips may be understood by separating the dominant frequencies in $P^s_{24}$:
\begin{align}
	P^s_{24}\equiv&\, C_-^s\, \cos[(1-A_e)\Delta]+ C_+^s \, \cos[(1+A_e)\Delta]+C_R^s \, \cos[(1-2R)\Delta]\;. \label{eq:sterileosc}
\end{align}
Here, the coefficients $C_{-}^{s}$, $C_{+}^{s}$ and $C_{R}^{s}$ are smoothly varying (non-oscillating) functions of $R$, $A_e$ and $\theta_{23}$, that regulate the amplitudes of peaks and dips, but do not affect their positions.
Examining each of the terms in eq.~(\ref{eq:sterileosc}), we can account for the oscillatory behaviors of the sterile neutrino contributions:
\begin{itemize}
	\item The first term in eq.~(\ref{eq:sterileosc}) oscillates as $\cos[(1-A_e)\Delta]$. 
	This term will contribute maximally at $(1-A_e)\Delta = n \pi$ i.e. at the dips of the  $3\nu$ contribution. This term will therefore modify the probability near the dips of $P_{\mu e}$, and hence will affect the determination of $\theta_{23}$ if sterile neutrinos are present.
	
	\item The second term in eq.~(\ref{eq:sterileosc}), which oscillates as $\cos[(1+A_e)\Delta]$, is not present in the $3\nu$ sector. However, note that the numerical value of this frequency for the $\nu$ channel is half of the leading order $3\nu$ frequency for the $\bar\nu$ channel, and vice versa ($\nu \leftrightarrow \bar\nu$).
	
	\item Exploring the final term of eq.~(\ref{eq:sterileosc}), we expect the sterile-induced peaks and dips to be at the extrema of $\cos[(1-2R)\Delta]$. Note that this frequency has no matter dependence. For $R>1$, this is the term which would induce oscillations at an energy higher than in the $3\nu$ case.
	The sign of this contribution would depend on the sign of $C_R^s  \sin(\delta_{24}^\prime) $. We can write
	\begin{equation}
		C_R^s=-\frac{1}{2} R \left( \frac{1}{A_e+2 R} +\frac{A_e+6 R-6}{\left(2 R-A_e\right) \left(A_e+2 R-2\right)} \right)\;. \label{eq:CSR}
	\end{equation} 
For $E_\nu =0.5-10$ GeV, with our parameter choices both $C_R^s$ and $\sin(\delta_{24}^\prime)$ are negative.
Therefore, the net sterile contribution peaks at $\cos [(1-2R)\Delta]=1$ and dips at $\cos [(1-2R)\Delta]=-1$, as can be seen in figure~\ref{fig:comparison}. 
Indeed, we can even calculate the positions of peaks and dips induced by sterile neutrinos. For example, the first two peaks and dips in the N-Ns-$\nu$ scenario (Top left in figure~\ref{fig:comparison}) can be seen to be at $(1-2R)\Delta= 2\pi, \;4\pi$ and $(1-2R)\Delta=\pi,\; 3\pi$ respectively:
\begin{align}
		E_\nu^\text{peak} \simeq & \; 3.55 \text{ GeV},\; 1.77 \text{ GeV}\;, \qquad
	E_\nu^\text{dip} \simeq \; 7.1 \text{ GeV},\; 2.37 \text{ GeV}\;.
\end{align}
Note that, though the peak and dip position are independent of matter effects, their amplitudes can have substantial  matter dependence, as can be seen from eq.~(\ref{eq:CSR}).
\end{itemize}
While the above discussion has been for N-Ns mass ordering in the neutrino sector, a similar analytic understanding for sterile peaks and dips may also be obtained for all the remaining mass ordering scenarios, viz. N-Is, I-Ns and I-Is, and also for antineutrinos.


In the next section, we will explore the sensitivity of DUNE to sterile mass ordering. The dependence of the sensitivity on $\Delta m^2_{41}$ can be explained using our analytic expressions obtained in this section.
\section{Sensitivity to sterile mass ordering at DUNE}
\label{sec:results}

DUNE (Deep Underground Neutrino Experiment) is an upcoming long-baseline experiment in the USA. It will consist of a neutrino source facility located at Fermilab and a far detector located at the Sanford Underground Research Facility in South Dakota, and thus will have a baseline of 1300 km. The primary aim of DUNE is to probe all the three unknowns in the $3\nu$ oscillation sector, viz., the leptonic CP violation, the neutrino mass ordering and the octant of $\theta_{23}$.
The accelerator at Fermilab will generate a proton beam of energy 80–120 GeV at 1.2–2.4 MW which will finally produce a neutrino beam of a wide energy range 0.5–8.0 GeV. The far detector will consist of four identical 10 kt LArTPC (Liquid Argon Time Projection Chamber) detectors with a total fiducial mass of 40 kt. We have used the General Long Baseline Experiment Simulator (GLoBES) package~\cite{Huber:2004ka,Huber:2007ji} to simulate the DUNE data.
The detector-related specifications used in this study are listed in table~\ref{dunedetails}. The neutrino oscillation parameter values used here are given in table~\ref{tab:steriletable}.
\begin{table}[h]
	\centering
	\begin{tabular}{ |l| l l |ll| }
		\hline
		Detector details & \multicolumn{2}{c |}{Normalization error} & \multicolumn{2}{c |}{\centering Energy calibration error} \\
		\cline{2-5}
		& Signal &\centering Background      & Signal & \quad Background                              
		\\ 
		\hline
		Baseline = 1300 km 
		&&&&\\
		Runtime (yr) = 3.5 $\nu$ + 3.5 $\bar \nu$ 
		& $\nu_e : 5\%$  &\quad $\nu_e : 10\%$ &   $\nu_e : 5\%$ &\qquad $\nu_e : 5\%$\\
		40 kton, LArTPC & & &&\\
		$\varepsilon_{app}=80\%$, $\varepsilon_{dis}=85\%$
		& &&&\\
		$R_e=0.15/\sqrt{E_\nu (\text{GeV})}$ , &$\nu_\mu : 5\%$ & \quad $\nu_\mu : 10\%$ & $\nu_\mu : 5\%$ &\qquad $\nu_\mu : 5\%$\\
		$R_\mu=0.20/\sqrt{E_\nu (\text{GeV})}$ &&&&\\
		\hline
	\end{tabular}
	\caption{\label{dunedetails} 
		Details of detector configurations, efficiencies, resolutions, and systematic uncertainties for DUNE. Here, $\varepsilon_{app}$ and $\varepsilon_{dis}$ are signal efficiencies for $\nu_{e}^{CC}$ and $\nu_{\mu}^{CC}$ respectively. Also, $R_e$ and $R_\mu$ are energy resolutions for $\nu_{e}^{CC}$ and $\nu_{\mu}^{CC}$ events respectively. One year of runtime corresponds to $1.47 \times 10^{21}$ POT (protons on target).}
\end{table}

\subsection{Analysis procedure}

We simulate the data by using the input (``true'') values of the parameters as given in table~\ref{tab:steriletable}, and try to fit the data with alternative (``test'') values of these parameters, corresponding to the opposite sterile mass ordering.
The quantity $\Delta \chi^2_{\text{SMO}}$ that quantifies the sensitivity of DUNE to sterile mass ordering is defined as
\begin{equation}
	\Delta \chi^2_{\text{SMO}} \equiv \chi^2 (\text{test})- \chi^2 (\text{true})\;,
\end{equation}
where the value of $\chi^2$ is obtained using the GLoBES package~\cite{Huber:2004ka,Huber:2007ji}.
We further perform minimization of $\chi^2$(test) by varying over the fitting parameters to take care of the effects of their uncertainties. The range of variation of the neutrino oscillation parameters has been given in table~\ref{tab:steriletable}.

\begin{table}[t]
	\centering
	\begin{tabular}{| c | c | c | c| }
		\hline
		Sector&Parameter & Value & Variation range \\ \hline\hline
		\multirow{6}{*}{Active} &$\theta_{12}$ & $33.56^\circ$  & -- \\ \cline{2-4} 
		&$\theta_{13}$ & $8.46^\circ$  & -- \\  \cline{2-4} 
		&$ \theta_{23}$ & $45^\circ$  &$[40^\circ, 50^\circ]$ \\ \cline{2-4} 
		&$\delta_{13}$ & $-90^\circ$ & $\left[-180^{\circ},0^{\circ}\right]$ \\	\cline{2-4} 
		&$\Delta m^2_{21 }$ & $7.5 \times 10^{-5}$ eV$^2$  & -- \\ \cline{2-4} 
		&$\Delta m^2_{31 }$ & $2.5 \times 10^{-3}$ eV$^2$  & -- \\  \hline	\hline
		\multirow{5}{*}{Sterile}  &$\theta_{14}$ & $5^\circ$ & $[0^\circ-\theta_{14}^\text{max}]$ \\ \cline{2-4} 
		&$\theta_{24}$ & $10^\circ$  & $[0^\circ - 55^\circ]$ \\ \cline{2-4}
		&$\theta_{34}$ & $0$ & -- \\ \cline{2-4} 
		&$\delta_{24}$ & $0$ &$\left[-180^{\circ},180^{\circ}\right]$ \\ \cline{2-4} 
		&$\delta_{34}$ & $0$ & -- \\ \cline{2-4}
		& $\Delta m^2_{41}$& $\Delta m^2_{41}$(true) &$\Delta m^2_{41}$(true)  $\pm 15\%$ \\
		\hline
	\end{tabular}
	\caption{The simulated (true) values of parameters in the active and sterile sectors, and the variation ranges taken for their test values.
	}
	\label{tab:steriletable} 
\end{table}

Among the active neutrino mixing parameters, $\theta_{12}$ and $\Delta m^2_{21}$ are not expected to affect the identification of sterile mass ordering. Further, the values of $\theta_{13}$ and $\Delta m^2_{31}$ are known to high precision, so we do not vary over these four parameters. We also take the mass ordering in the active neutrino sector to be known. However, we vary over $\theta_{23}$ and $\delta_{13}$, which have large uncertainties.

In the sterile sector, we choose to restrict our analysis to $\theta_{34}\text{(test)}=0$, i.e. we do not vary over $\theta_{34}$ for the sake of practicality. This also makes the value of $\delta_{34}$ irrelevant, allowing us to focus on the dominant effects of $P^s_{24}$.
We then vary over the two mixing angles $\theta_{14}$ and $\theta_{24}$, the CP violating phase $\delta_{24}$, and the mass squared difference $\Delta m^2_{41}$, in the ranges shown in table~\ref{tab:steriletable}. For the range of $\theta_{24}$, we use a conservative upper bound based on constraints from MINOS and MINOS+~\cite{DUNE:2020fgq,MINOS:2017cae}. For $\theta_{14}$, we use a conservative upper bound based on the constraints from Daya Bay/Bugey-3~\cite{DUNE:2020fgq} as well as those from $\sin^2 \theta_{\mu e} \equiv \sin^2 2\theta_{14} \sin^2 \theta_{24}$ at MINOS and MINOS+~\cite{MINOS:2017cae}. Our variation range for $\Delta m^2_{41}$ considers a $\pm 15\%$ error in its measurement
\footnote{
	If DUNE observes sterile neutrinos with $|\Delta m^2_{41}| \lesssim O(10^{-2}) \text{ eV}^2$, then it would measure the active-sterile oscillation phase with a precision of
	\begin{equation}
		\frac{\delta\phi }{2\pi} = \frac{1}{2\pi}\; \frac{\Delta m^2_{41} L}{4 E_\nu} \; \frac{\delta E_\nu}{E_\nu} < 0.14
	\end{equation}
	for $E_\nu >2$ GeV, where we have taken the energy resolution $\delta E_\nu/E_\nu = 0.15/\sqrt{E_\nu (\text{GeV})}$. This is the same as the precision in $\Delta m^2_{41} $. Therefore, we can safely take the precision in $\Delta m^2_{41}$ to be better than $\pm 15\%$ in this low-$\Delta m^2_{41}$ range. As we will see later in this section, the sensitivity to sterile mass ordering is very low for $\Delta m^2_{41} \gtrsim 10^{-2} \text{ eV}^2$. Hence, the precision in this high-$\Delta m^2_{41}$ range will not matter in our analysis.
}.

\subsection{Dependence of the sensitivity on $|\Delta m^2_{41}|$}
\label{sec:delchidelm}
\begin{figure}[t]
	\centerline{\includegraphics[width=0.75\textwidth]{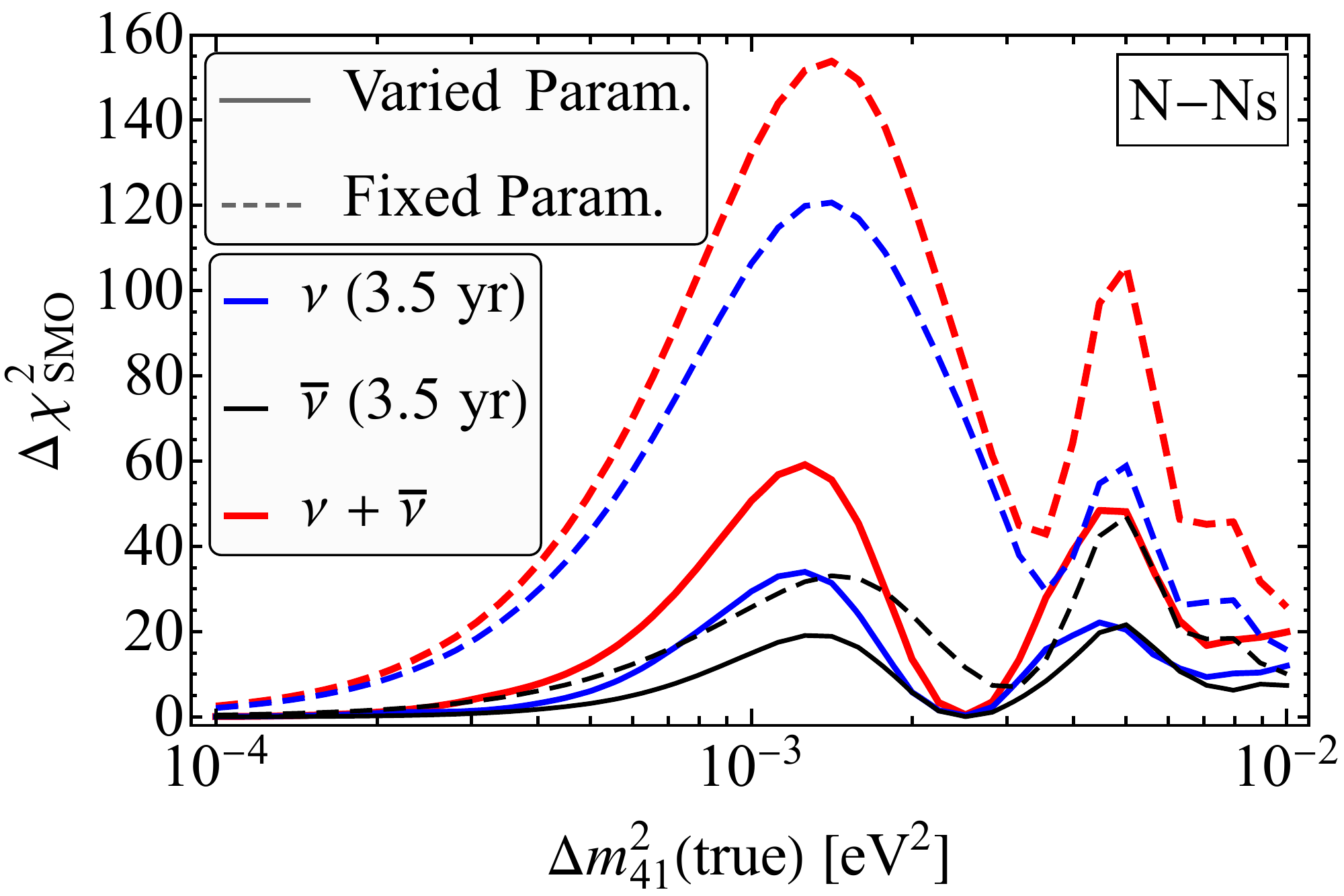}}
	\caption{Dependence of $\Delta \chi^2_{\text{SMO}}$ on $\Delta m^2_{41}$, for the N-Ns ($\Delta m^2_{31} >0, \Delta m^2_{41} >0$) scenario. The dashed curves are obtained by using fixed test values of neutrino mixing parameters. The solid curves are obtained by varying the test values in the ranges indicated in table~\ref{tab:steriletable}. The neutrino and the antineutrino runs are taken to be for 3.5 years each.}
	\label{fig:marg}
\end{figure}

The sensitivity to sterile mass ordering at DUNE calculated over a wide range of $|\Delta m^2_{41}|$ is shown in figure~\ref{fig:marg}, taking the N-Ns scenario ($\Delta m^2_{31}>0$, $\Delta m^2_{41}>0$). We show the results separately for the $3.5$ yr neutrino run, for the $3.5$ yr antineutrino run, and their combination. The results are presented for the cases when the test parameters are fixed, as well as when they are varied over their ranges specified in table~\ref{tab:steriletable}.

The following observations can be made from figure~\ref{fig:marg}:
\begin{itemize}
	\item In principle, DUNE has sensitivity to sterile mass ordering over the $|\Delta m^2_{41}|$ range of $ (10^{-4}-10^{-2}) \text{ eV}^2$. This is not surprising since one of the major aims of DUNE is to observe the mass ordering around $\Delta m^2_{31} \approx 2.5\times 10^{-3} \text{ eV}^2$. For $|\Delta m^2_{41}| < 10^{-4} \text{ eV}^2$, oscillations due to sterile neutrino would not develop for DUNE. For $|\Delta m^2_{41}| >10^{-2} \text{ eV}^2$, we expect a reduced sensitivity to sterile mass ordering due to multiple reasons, viz. the averaging out of sterile neutrino oscillations, the reduced effect on matter potential terms, and a reduced interference between the frequencies $\Delta m^2_{31}$ and $\Delta m^2_{41}$.
	
	\item The variation over the uncertainties of the neutrino mixing parameters decreases the sensitivity considerably --- almost by a factor of 3. In spite of this, it is observed that over a wide range of $|\Delta m^2_{41}|$ values, it is possible to have $\Delta \chi^2_{\text{SMO}} \gtrsim 25$ (i.e. a 5$\sigma$ identification of sterile mass ordering), when neutrino and antineutrino data are combined.
	
	\item We observe a dip in sensitivity to sterile mass ordering at $|\Delta m^2_{41}|  \approx |\Delta m^2_{31}|$, this is due to possible degeneracy between the sterile and the atmospheric mass squared difference, which makes it difficult to disentangle their contributions.
	
	\item The $\Delta \chi^2_{\text{SMO}}$ value in the neutrino (for fixed parameters as well as when they are varied) channel is considerably larger than that in the antineutrino channel. This is expected, since the cross sections for antineutrinos at $\sim$GeV energies are approximately half of the neutrino cross sections. However, note that for $\Delta m^2_{41} > \Delta m^2_{31}$, the sensitivity for the antineutrino channel increases significantly and becomes almost comparable to the neutrino channel sensitivity.
	The reason for this can be understood by inspecting the $P^s_{24}$ term in eq.~(\ref{eq:Ps24}) term that regulates the sterile contribution to $P_{\mu e}$. In the antineutrino channel for the N-Ns scenario, we have a possible resonant behavior at $R>1$, i.e. for $\Delta m^2_{41} > \Delta m^2_{31}$, leading to a enhanced change in the conversion probability. However, in the neutrino channel, no such resonances involving sterile neutrinos are possible for the N-Ns scenario and $R>1$.
\end{itemize}

The first two observations above will be seen to hold when we later discuss the other three mass ordering combinations, viz. N-Is, I-Ns, and I-Is.
The last two observations will be modified depending on the mass ordering combinations. This will be analyzed in the next section.

\subsection{Dependence of sensitivity on mass ordering combinations}
\label{subsec:results}
\begin{table}[t]
	\centering
			\begin{tabular}{|c|c|c|c|c|}
				\hline
				Active $\nu$ mass ordering & Sterile $\nu$ mass ordering & $\nu / \bar\nu$& sign of $A_e$& sign of $R$\\
				\hline
				\multirow{4}{*}{N} & \multirow{2}{*}{Ns} & $\nu$ &$+$&$+$\\\cline{3-5}
				& & $\bar\nu$ &$-$&$+$\\ \cline{2-5}
				& \multirow{2}{*}{Is}& $\nu$ &$+$&$-$\\ \cline{3-5}
				& &  $\bar\nu$&$-$&$-$ \\ \hline
				\multirow{4}{*}{I} & \multirow{2}{*}{Ns} & $\nu$ &$-$&$-$\\\cline{3-5}
				& & $\bar\nu$ &$+$&$-$\\ \cline{2-5}
				& \multirow{2}{*}{Is}& $\nu$&$-$&$+$ \\ \cline{3-5}
				& &  $\bar\nu$ &$+$&$+$\\ \hline
			\end{tabular}
		\caption{The signs of $A_e$ and $R$ for all mass ordering combinations, for neutrinos and antineutrinos.}
		\label{tab:plusminus}
	\end{table}
\begin{figure}[t]
	\includegraphics[width=0.48\textwidth]{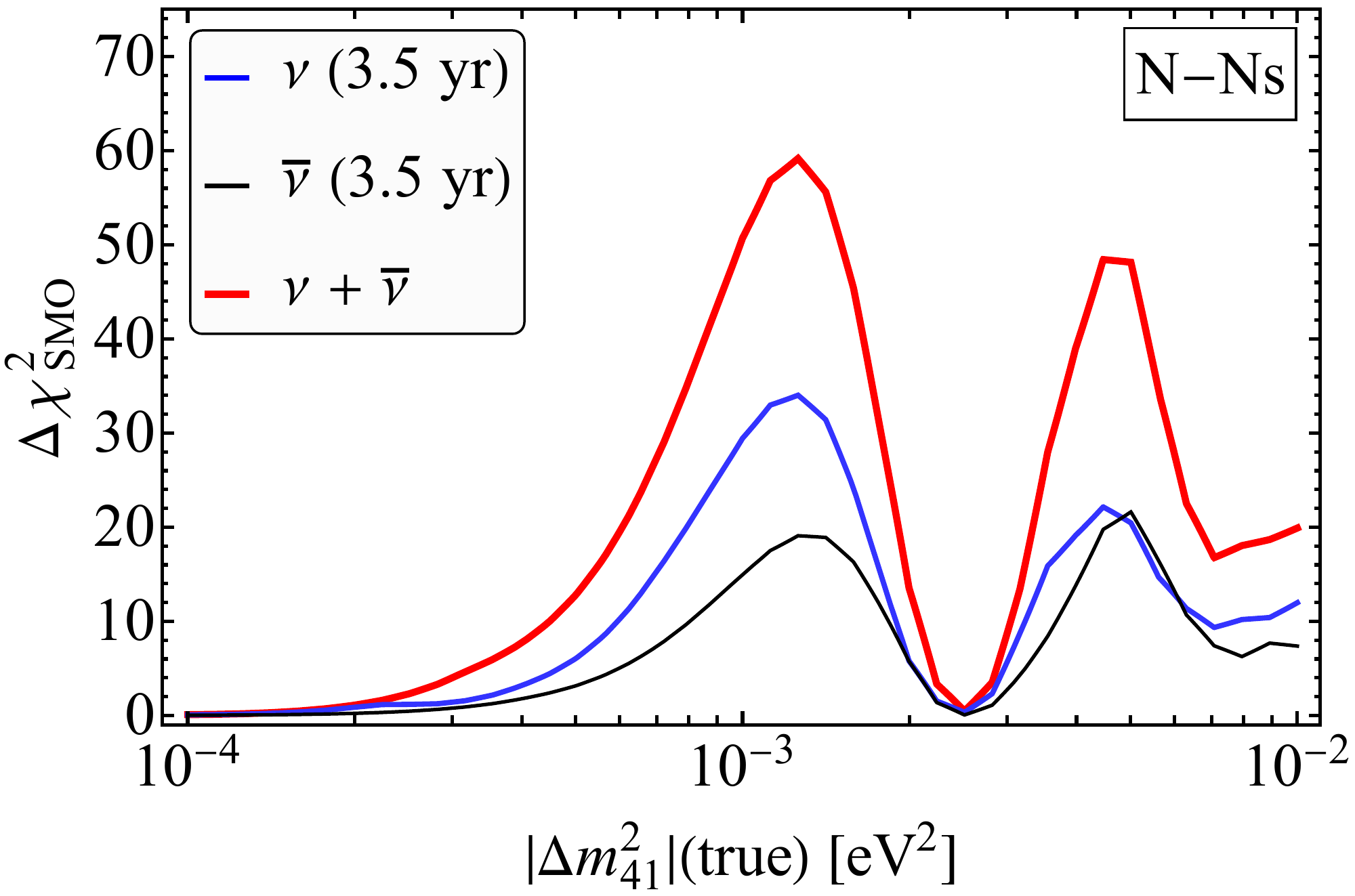}
	\includegraphics[width=0.48\textwidth]{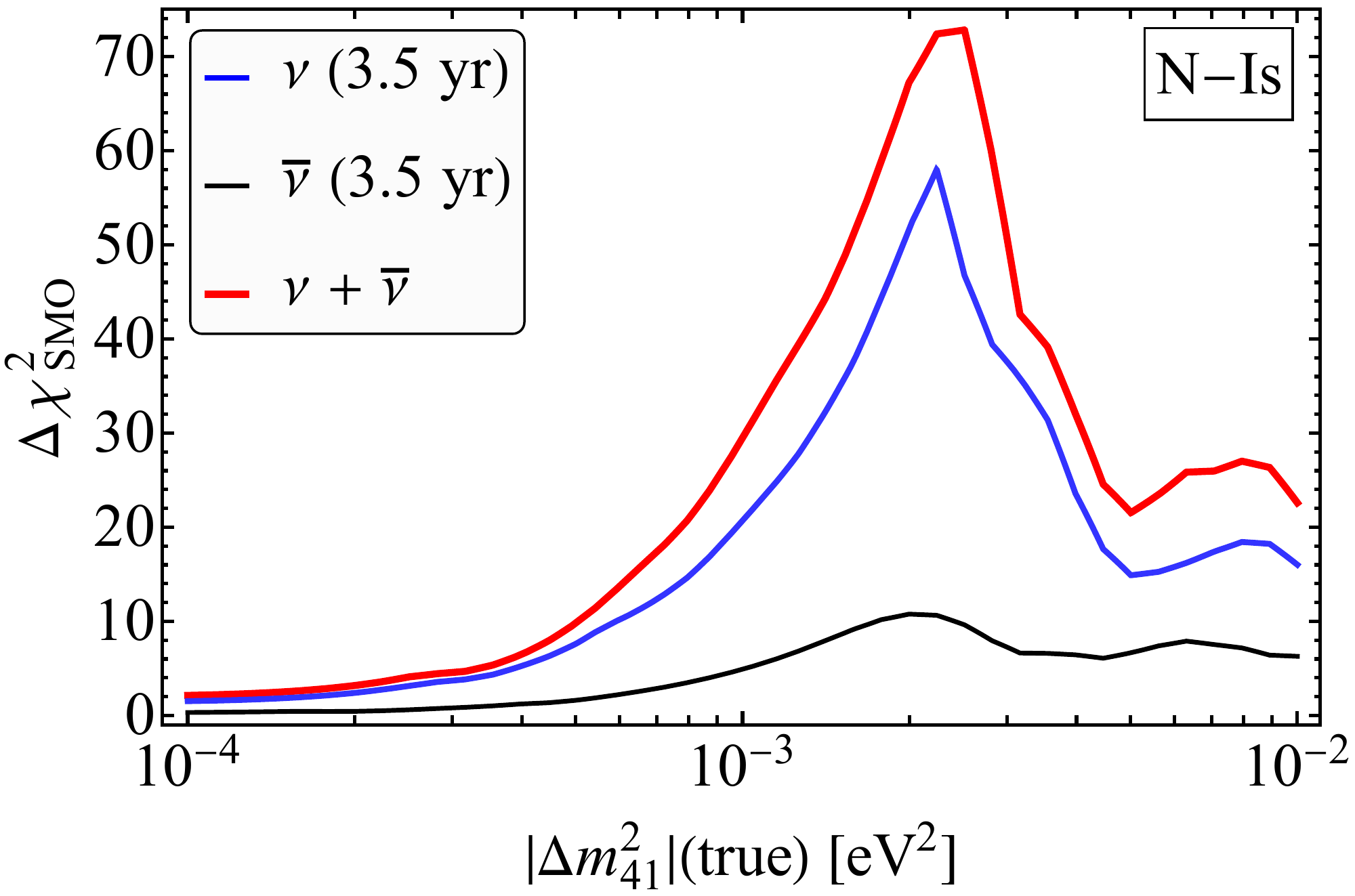}\\
	\vspace{0.1pt}\\
	\includegraphics[width=0.48\textwidth]{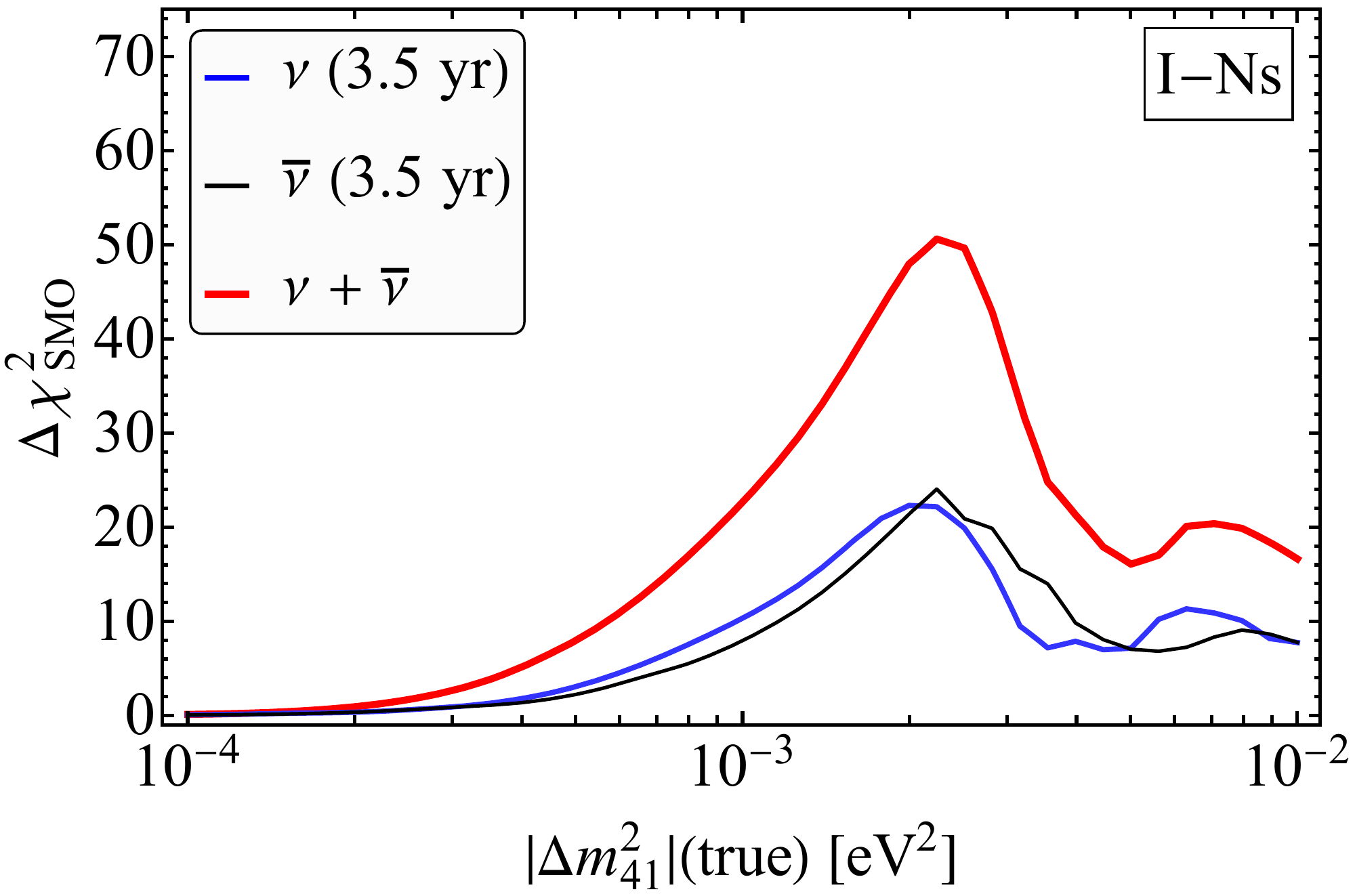}
	\includegraphics[width=0.48\textwidth]{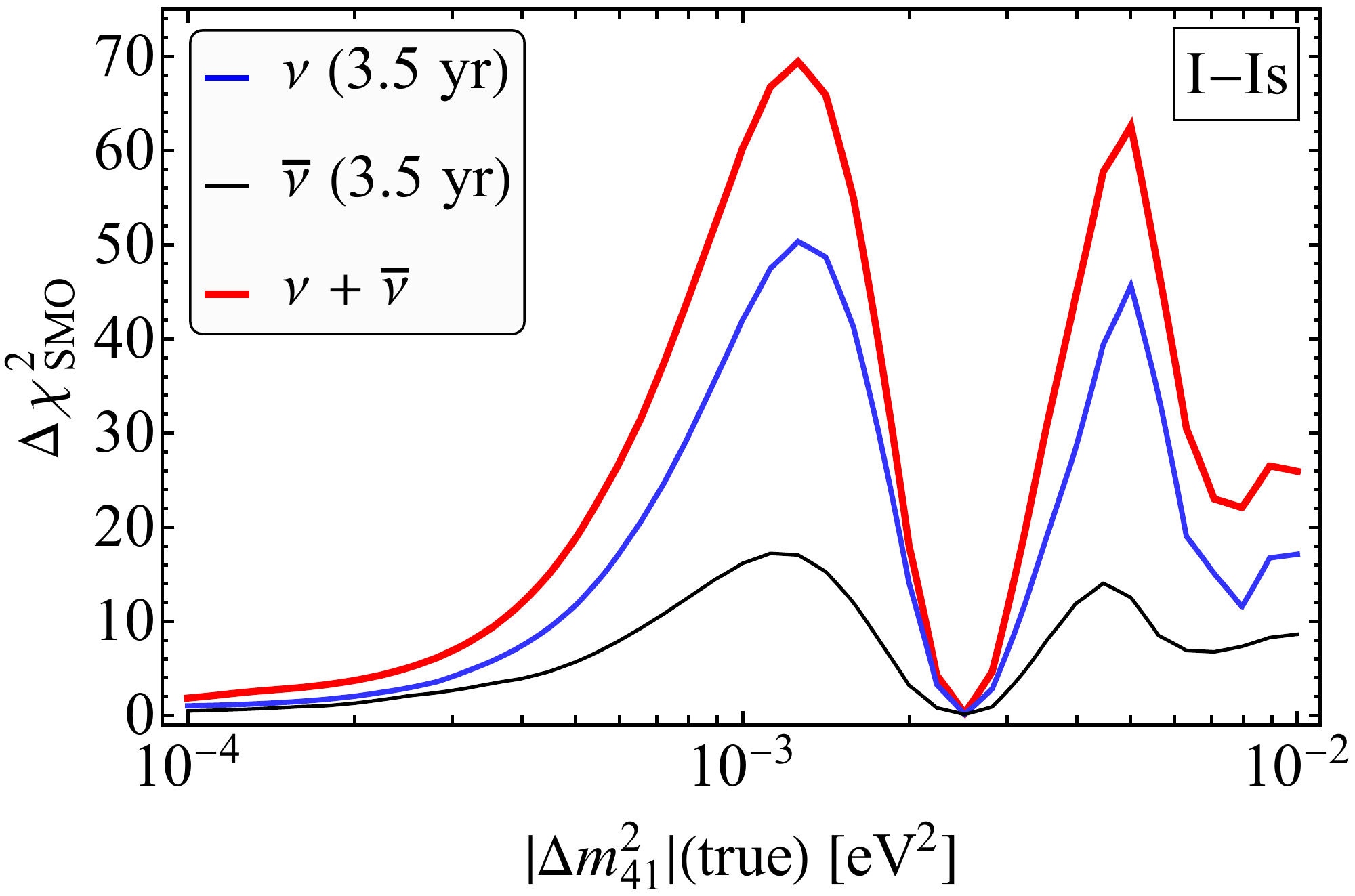}
	\caption{Same as the figure~\ref{fig:marg}, but for all four mass ordering combinations (N-Ns, N-Is, I-Ns, I-Is). We show only the results with test parameters varied over the range indicated in Table~\ref{tab:steriletable}.}
	
	\label{fig:all4results}
\end{figure}

In the last section, we explored the details of sensitivity to $|\Delta m^2_{41}|$ for the N-Ns scenario. In this section, we will explore this dependence for other scenarios, viz.  N-Is, I-Ns, and I-Is, and draw comparisons among them. These differences will be explained by our analytic approximations obtained in section~\ref{sec:analytical}.
In figure~\ref{fig:all4results}, we show the sensitivity to sterile mass ordering for all the above mentioned combinations, for the $3.5$ yr neutrino run, $3.5$ yr antineutrino run, and their combined statistics.
Note that now we only show the results where the neutrino oscillation parameters are varied over their uncertainties. 

Before we further discuss the effects of matter potential in the sterile contribution to $P_{\mu e}$, we first document the signs of $A_e$ and $R$ for all possible mass ordering combinations, for both neutrinos and antineutrinos, in table~\ref{tab:plusminus}. We see that the 8 different scenarios can be classified into 4 distinct sets corresponding to the sign of $A_e$ and $R$, viz. $(+,+),\;(-,+),\;(+,-),\;(-,-)$. We expect that the behavior due to matter effect and sterile term will be uniform within these four independent sets.
In Fig.~\ref{fig:all4results} we observe that:
\begin{itemize}
	\item Even taking into account the dilution in sensitivity due to variation over most of the test parameters, DUNE remains sensitive to sterile mass ordering for all mass ordering combinations, for $|\Delta m^2_{41}| \in (10^{-4}, 10^{-2}) \text{ eV}^2$.
	
	\item For N-Ns and I-Is scenarios, we observe a dip in sensitivity  at $|\Delta m^2_{41}|  \approx |\Delta m^2_{31}| $. As noted in section~\ref{sec:delchidelm}, this is due to the degeneracy between $\Delta m^2_{31}$ and $\Delta m^2_{41}$. This degeneracy does not occur in the scenarios N-Is and I-Ns, since the mass squared differences have opposite signs. As a result, the sharp dips present in the earlier two scenarios are absent in these two.
	
	\item  We observe higher sensitivity in the neutrino channel for N-Ns-$\nu$, N-Is-$\nu$ and in the antineutrino channel, for I-Ns-$\bar\nu$, I-Is-$\bar\nu$. This is due to the enhancement from the $\sin [(A_e-1)\Delta] / (A_e-1)$ factor in the sterile contribution to $P_{\mu e}$. One can see from table~\ref{tab:plusminus} that for the above mentioned combinations $A_e$ is positive, leading to the enhancement. This is thus due to the interplay between active neutrino  mass ordering and matter effects.
	
	\item Similarly, we observe increased sensitivity for N-Is-$\nu$, I-Is-$\nu$ in the neutrino channel and for N-Ns-$\bar\nu$, I-Ns-$\bar\nu$ in the antineutrino channel. This may be explained by the interplay between the two new parameters in the Hamiltonian, $A_n (=-A_e/2)$ and $R$, that become relevant for sterile neutrino propagation in matter. The resulting $\sin [(R+A_e/2)\Delta]/(R+A_e/2)$ factor in the sterile contribution to $P_{24}^s$, as shown in eq.~(\ref{eq:Ps24}) leads to an enhancement in the $(+,-)$ and $(-,+)$ scenarios in table~\ref{tab:plusminus}.
\end{itemize}

In the next section, we shall explore  in more detail the possible enhancement in the sensitivity to sterile mass ordering due to the sterile contribution to $P_{\mu e}$ giving rise to resonance-like behaviors, that appear in certain mass ordering combinations in matter.

\subsection{Interplay between $\Delta m^2_{41}$ and matter effects}

\begin{table}[t]
	\centering
	\begin{tabular}{|c|c|c|c|c|c|}
		\hline
		\multirow{2}{*}{sign of $A_e$}& \multirow{2}{*}{sign of $R$} & \multicolumn{2}{c|}{Combinations} &\multirow{2}{*}{$|R|<1$} &\multirow{2}{*}{$|R|>1$}\\
		\cline{3-4}
		&&$\qquad$$\nu$$\qquad$&$\qquad$$\bar\nu$$\qquad$&&\\\hline
		$+$&$+$& N-Ns-$\nu$ & I-Is-$\bar\nu$ & $\checkmark$ & --- \\ \hline
		$-$&$+$& I-Is-$\nu$ & N-Ns-$\bar\nu$ & $\checkmark$ & $\checkmark\checkmark$ \\ \hline
		$+$&$-$& N-Is-$\nu$ & I-Ns-$\bar\nu$ & $\checkmark\checkmark$ & ---\\ \hline
		$-$&$-$&  I-Ns-$\nu$ & N-Is-$\bar\nu$ & --- & --- \\ \hline
	\end{tabular}
	\caption{Modifications in the probabilities $P(\nu_\mu \to \nu_e)$ and $P(\bar\nu_\mu \to \bar \nu_e)$ due to the interplay between $\Delta m^2_{41}$ and matter effect. Here dash (`---') denotes the absence of significant enhancement due to matter effects. The single tick (`$\checkmark$') denotes a small enhancement and  double ticks (`$\checkmark\checkmark$') denote a large enhancement due to possible resonance-like behaviors.}
	\label{tab:sensitivity}
\end{table}

For the parameter choices in table~\ref{tab:steriletable}, the relevant sterile contribution in the conversion channel may be expressed as
\begin{equation}
	P_{\mu e} (\text{sterile})=4 \, s_{13} \,  s_{14}  \, s_{24} \, s_{23} \, \frac{\sin \left[\left(A_e-1\right)\Delta\right]}{A_e-1} \sin( \delta_{24}^\prime) P_{24}^{s}\;.
\end{equation}
Here, the $P_{24}^{s}$ term [eq.~(\ref{eq:Ps24})] is
\begin{align}
	\label{eq:Ps24result}
	P_{24}^s=&\;R\Big[\tfrac{1}{2} A_e c_{23}^2 +(R-1) \left(s_{23}^2+1\right)\Big]\frac{\sin   \left[\left(R-1+\frac{A_e}{2}\right)\Delta\right] }{ R-1+\frac{A_e}{2} }\frac{\sin\left[ \left(R-\frac{A_e}{2}\right) \Delta\right] }{R-\frac{A_e}{2}}\nonumber\\
	&+c_{23}^2 R \sin \left[\left(R-1-\tfrac{A_e}{2}\right)\Delta\right] \frac{\sin \left[\left(R+\frac{A_e}{2}\right) \Delta\right] }{R+\frac{A_e}{2}}\;.
\end{align}
The normalized effective matter potential is $A_e\approx 0.09 \times (E_\nu \text{ in GeV})$ for DUNE. Eq.~(\ref{eq:Ps24result}) indicates that resonance-like behaviors would appear when   $A_e\approx2(1-R)$ or $A_e\approx\pm2R$. Resonance due to the first condition would appear only for one of the possible signs of $R$. Even though the second condition may be satisfied for both signs of $R$, their coefficients in eq.~(\ref{eq:Ps24result}) are different. This leads to a different value of $P^s_{24}$ for different signs of `$R$'. Therefore, such resonance like behavior may be expected to lead to a higher sensitivity to the sterile mass ordering.

 The occurrence and strength of the enhancement in sensitivity due to such resonance-like behaviors is predicted in table~\ref{tab:sensitivity}.
We can now explain the following features of figure~\ref{fig:all4results}: 

\begin{itemize}
	\item The cross section of neutrinos of GeV energy is higher than those of antineutrinos by a factor of about two. Therefore, in the absence of matter effects, one would expect the sensitivity in the neutrino channel to be about twice that in the antineutrino channel. Major deviations from this naive expectation occur in the following scenarios:
	\begin{itemize}
		\item[1.] The relative sensitivity in the antineutrino channel is enhanced in the scenarios (i) N-Ns with $|R|>1$ and (ii) I-Ns with $|R|<1$. In the first scenario, there is enhancement in the antineutrino channel near $ A_e =2(1-R)$, which is absent in the neutrino channel. Similarly, in the second scenario, the enhancement is present only in the antineutrino channel near $A_e=-2R$.
		\item[2.] The relative sensitivity in the neutrino channel is enhanced in the scenarios (i)~I-Is with $|R|>1$ and (ii) N-Is with $|R|<1$. In the first scenario, there is enhancement in the neutrino channel near $ A_e =2(1-R)$, which is absent in the neutrino channel. Similarly, in the second scenario, the enhancement is present only in the neutrino channel near $A_e=-2R$.
	\end{itemize}

	\item In the $|R|>1$ regime, the sensitivity for the N-Ns and I-Is combinations  is larger than that for the N-Is and I-Ns mass orderings. This is due to the enhancement in sensitivity for the N-Ns-$\bar\nu$ and I-Is-$\nu$ probabilities, respectively. Both  these combination belong to the $(-,+)$ set in table~\ref{tab:sensitivity}. For this set, enhancement is expected due to the resonance-like condition near $ A_e =2(1-R)$.  Even among these two mass orderings, the sensitivity for I-Is is more since the enhancement is in the neutrino channel. Such an enhancement does not occur for N-Is and I-Ns mass orderings in $\nu$ or $\bar\nu$ channels.
	
	\item In the $|R|<1$ regime, the sensitivity in the N-Is mass ordering is high due to a resonance-like behavior near $A_e=-2R$ in the neutrino channel. A similar enhancement occurs for the I-Ns mass ordering, albeit in the antineutrino channel, so the overall enhancement is not as pronounced. Further, for both N-Ns and I-Is mass ordering scenarios, enhancements due resonance-like behavior can occur in both the $\nu$ and $\bar\nu$ channels as pointed out in table~\ref{tab:sensitivity}.
	
	\item For the I-Ns mass ordering, the sensitivity in the neutrino channel is small, leading to an overall low sensitivity. This is due to the lack of any enhancement in the neutrino channel for this mass ordering. 
\end{itemize}
This demonstrates the power of our analytic approximation for the conversion probability, and the utility of representing it in the $\sin(x)/x$ form as shown in section~\ref{sec:analytical}.

\section{Conclusions}
\label{sec:concl}

In this paper, we analytically calculate the conversion probability $P_{\mu e}$ in the presence of sterile neutrinos, with exact dependence on $\Delta m^2_{41}$ and with explicit dependence on matter potential.
The probability is expressed as a perturbative expansion in the small parameters $\alpha$, $s_{13}$, $s_{14}$, $s_{24}$ and $s_{34}$.
We show that the terms involving $s_{24}$ and $s_{34}$ can be explicitly separated, with the latter term contributing only in the presence of matter, due to the neutral-current forward-scattering of active neutrinos.
Further, we rearrange the probability expression in terms of the $\sin (x)/x$ form and show that the dependence on CP-violating angles in the sterile sector ($\delta_{24}$ and $\delta_{34}$) can be separated. This form encapsulates the resonance-like behaviors occurring when matter potentials and $\Delta m^2_{41}$ satisfy specific relationships.

To bring out the power of our formalism, we first show that our analytic expression can accurately predict the  positions and amplitudes of sterile induced oscillations at a long-baseline experiment like DUNE. We further focus on the identification of sterile mass ordering, i.e. the sign of $\Delta m^2_{41}$, at DUNE, and motivate that such an identification is possible for $\Delta m^2_{41} \in (10^{-4}-10^{-2}) \text{ eV}^2$ for a wide choice of neutrino-mixing parameter values. Note that this mass-squared range overlaps the parameter space which can address the tension between T2K and NOvA data. Since the mass-squared scales $\Delta m^2_{31}$ and $\Delta m^2_{41}$ are comparable to each other in this range, it is important to calculate the explicit contributions of sterile oscillations in matter. Our analytic expressions, therefore, are particularly crucial for probing the effects of sterile neutrinos for such scenarios.

We numerically calculate the sensitivity of DUNE to sterile mass ordering for all the mass ordering combinations in the active and sterile sector. We find that this sensitivity can indeed be significant in the range $\Delta m^2_{41} \sim (10^{-4}-10^{-2}) \text{ eV}^2$. This is expected, since DUNE is designed to probe the parameter range around such values. This sensitivity is observed to have intricate dependence on the actual value of $\Delta m^2_{41}$, which, however can be clearly understood by the analytic approximations calculated in this paper. In particular, these approximations can explain the relative sensitivities in neutrino and the antineutrino channel, in the various mass ordering scenarios, in terms of the resonance-like behaviors. The non-trivial effects of the complex interplay between $\Delta m^2_{41}$ and the matter effects can thus be clearly understood.

Although our analysis has been focused on DUNE and the identification of sterile mass ordering therein, the expressions for $P_{\mu e}$ that we have calculated would be valid for all current and upcoming long-baseline experiments. In general, they would be valid as long as the matter densities neutrinos propagate through may be approximated by a single line-averaged density. Thus, even for atmospheric neutrinos that do not pass through the core, our expressions would serve as a good approximation which is valid over a wide range of~$\Delta m^2_{41}$.


\acknowledgments
The authors would like to acknowledge WHEPP-XVI for their hospitality and arrangements, where initial discussions and work took place. D.S.C. and A.D. acknowledge support from the Department of Atomic Energy (DAE), Government of India, under Project Identification No. RTI4002. D.P. is thankful for the support of FAPESP funding Grant 2014/19164-6 and 2020/04261-7. M.M.D. would like to acknowledge the support of the DST SERB grant EMR/2017/001436. D.S.C. thanks K.~Ghadiali for his help regarding the numerical packages.


\providecommand{\href}[2]{#2}\begingroup\raggedright\endgroup


\begin{thebibliography}{10}
	
	\bibitem{SajjadAthar:2021prg}
	M.~Sajjad~Athar et~al., \emph{{Status and perspectives of neutrino physics}},
	\href{https://doi.org/10.1016/j.ppnp.2022.103947}{\emph{Prog. Part. Nucl.
			Phys.} {\bfseries 124} (2022) 103947}
	[\href{https://arxiv.org/abs/2111.07586}{{\tt arXiv:2111.07586}}].
	
	\bibitem{Workman:2022ynf}
	{\scshape Particle Data Group} collaboration, \emph{{Review of Particle
			Physics}}, \href{https://doi.org/10.1093/ptep/ptac097}{\emph{PTEP} {\bfseries
			2022} (2022) 083C01}.
	
	\bibitem{Esteban:2020cvm}
	I.~Esteban, M.C.~Gonzalez-Garcia, M.~Maltoni, T.~Schwetz and A.~Zhou,
	\emph{{The fate of hints: updated global analysis of three-flavor neutrino
			oscillations}}, \href{https://doi.org/10.1007/JHEP09(2020)178}{\emph{JHEP}
		{\bfseries 09} (2020) 178} [\href{https://arxiv.org/abs/2007.14792}{{\tt
			arXiv:2007.14792}}].
	
	\bibitem{nufit}
	NuFIT v5.1 (2021), \url{http://www.nu-fit.org}.
	
	\bibitem{deSalas:2020pgw}
	P.F.~de~Salas, D.V.~Forero, S.~Gariazzo, P.~Mart\'\i{}nez-Mirav\'e, O.~Mena,
	C.A.~Ternes et~al., \emph{{2020 global reassessment of the neutrino
			oscillation picture}},
	\href{https://doi.org/10.1007/JHEP02(2021)071}{\emph{JHEP} {\bfseries 02}
		(2021) 071} [\href{https://arxiv.org/abs/2006.11237}{{\tt
			arXiv:2006.11237}}].
	
	\bibitem{Capozzi:2021fjo}
	F.~Capozzi, E.~Di~Valentino, E.~Lisi, A.~Marrone, A.~Melchiorri and A.~Palazzo,
	\emph{{Unfinished fabric of the three neutrino paradigm}},
	\href{https://doi.org/10.1103/PhysRevD.104.083031}{\emph{Phys. Rev. D}
		{\bfseries 104} (2021) 083031} [\href{https://arxiv.org/abs/2107.00532}{{\tt
			arXiv:2107.00532}}].
	
	\bibitem{Hyper-KamiokandeProto-:2015xww}
	{\scshape Hyper-Kamiokande Proto-} collaboration, \emph{{Physics potential of a
			long-baseline neutrino oscillation experiment using a J-PARC neutrino beam
			and Hyper-Kamiokande}},
	\href{https://doi.org/10.1093/ptep/ptv061}{\emph{PTEP} {\bfseries 2015}
		(2015) 053C02} [\href{https://arxiv.org/abs/1502.05199}{{\tt
			arXiv:1502.05199}}].
	
	\bibitem{ICAL:2015stm}
	{\scshape ICAL} collaboration, \emph{{Physics Potential of the ICAL detector at
			the India-based Neutrino Observatory (INO)}},
	\href{https://doi.org/10.1007/s12043-017-1373-4}{\emph{Pramana} {\bfseries
			88} (2017) 79} [\href{https://arxiv.org/abs/1505.07380}{{\tt
			arXiv:1505.07380}}].
	
	\bibitem{JUNO:2015zny}
	{\scshape JUNO} collaboration, \emph{{Neutrino Physics with JUNO}},
	\href{https://doi.org/10.1088/0954-3899/43/3/030401}{\emph{J. Phys. G}
		{\bfseries 43} (2016) 030401} [\href{https://arxiv.org/abs/1507.05613}{{\tt
			arXiv:1507.05613}}].
	
	\bibitem{DUNE:2020ypp}
	{\scshape DUNE} collaboration, \emph{{Deep Underground Neutrino Experiment
			(DUNE), Far Detector Technical Design Report, Volume II: DUNE Physics}},
	\href{https://arxiv.org/abs/2002.03005}{{\tt arXiv:2002.03005}}.
	
	\bibitem{DUNE:2020jqi}
	{\scshape DUNE} collaboration, \emph{{Long-baseline neutrino oscillation
			physics potential of the DUNE experiment}},
	\href{https://doi.org/10.1140/epjc/s10052-020-08456-z}{\emph{Eur. Phys. J. C}
		{\bfseries 80} (2020) 978} [\href{https://arxiv.org/abs/2006.16043}{{\tt
			arXiv:2006.16043}}].
	
	\bibitem{ALEPH:2005ab}
	{\scshape ALEPH, DELPHI, L3, OPAL, SLD, LEP Electroweak Working Group, SLD
		Electroweak Group, SLD Heavy Flavour Group} collaboration, \emph{{Precision
			electroweak measurements on the $Z$ resonance}},
	\href{https://doi.org/10.1016/j.physrep.2005.12.006}{\emph{Phys. Rept.}
		{\bfseries 427} (2006) 257} [\href{https://arxiv.org/abs/hep-ex/0509008}{{\tt
			hep-ex/0509008}}].
	
	\bibitem{doi:10.1142/5024}
	R.N.~Mohapatra and P.B.~Pal, \emph{Massive Neutrinos in Physics and
		Astrophysics}, WORLD SCIENTIFIC, 3rd~ed. (2004),
	\href{https://doi.org/10.1142/5024}{10.1142/5024}.
	
	\bibitem{giunti2007fundamentals}
	C.~Giunti and C.~Kim, \emph{Fundamentals of Neutrino Physics and Astrophysics},
	OUP Oxford (2007).
	
	\bibitem{King:2014nza}
	S.F.~King, A.~Merle, S.~Morisi, Y.~Shimizu and M.~Tanimoto, \emph{{Neutrino
			Mass and Mixing: from Theory to Experiment}},
	\href{https://doi.org/10.1088/1367-2630/16/4/045018}{\emph{New J. Phys.}
		{\bfseries 16} (2014) 045018} [\href{https://arxiv.org/abs/1402.4271}{{\tt
			arXiv:1402.4271}}].
	
	\bibitem{zuber2020neutrino}
	K.~Zuber, \emph{Neutrino physics}, Taylor \& Francis (2020).
	
	\bibitem{Dasgupta:2021ies}
	B.~Dasgupta and J.~Kopp, \emph{{Sterile Neutrinos}},
	\href{https://doi.org/10.1016/j.physrep.2021.06.002}{\emph{Phys. Rept.}
		{\bfseries 928} (2021) 1} [\href{https://arxiv.org/abs/2106.05913}{{\tt
			arXiv:2106.05913}}].
	
	\bibitem{Athanassopoulos:1995iw}
	{\scshape LSND} collaboration, \emph{{Candidate events in a search for
			anti-muon-neutrino ---\ensuremath{>} anti-electron-neutrino oscillations}},
	\href{https://doi.org/10.1103/PhysRevLett.75.2650}{\emph{Phys. Rev. Lett.}
		{\bfseries 75} (1995) 2650}
	[\href{https://arxiv.org/abs/nucl-ex/9504002}{{\tt nucl-ex/9504002}}].
	
	\bibitem{Aguilar:2001ty}
	{\scshape LSND} collaboration, \emph{{Evidence for neutrino oscillations from
			the observation of $\bar{\nu}_e$ appearance in a $\bar{\nu}_\mu$ beam}},
	\href{https://doi.org/10.1103/PhysRevD.64.112007}{\emph{Phys. Rev. D}
		{\bfseries 64} (2001) 112007}
	[\href{https://arxiv.org/abs/hep-ex/0104049}{{\tt hep-ex/0104049}}].
	
	\bibitem{Gariazzo:2018mwd}
	S.~Gariazzo, C.~Giunti, M.~Laveder and Y.F.~Li, \emph{{Model-independent
			$\bar\nu_{e}$ short-baseline oscillations from reactor spectral ratios}},
	\href{https://doi.org/10.1016/j.physletb.2018.04.057}{\emph{Phys. Lett. B}
		{\bfseries 782} (2018) 13} [\href{https://arxiv.org/abs/1801.06467}{{\tt
			arXiv:1801.06467}}].
	
	\bibitem{MiniBooNE:2018esg}
	{\scshape MiniBooNE} collaboration, \emph{{Significant Excess of ElectronLike
			Events in the MiniBooNE Short-Baseline Neutrino Experiment}},
	\href{https://doi.org/10.1103/PhysRevLett.121.221801}{\emph{Phys. Rev. Lett.}
		{\bfseries 121} (2018) 221801} [\href{https://arxiv.org/abs/1805.12028}{{\tt
			arXiv:1805.12028}}].
	
	\bibitem{PhysRevD.103.052002}
	{\scshape MiniBooNE Collaboration} collaboration, \emph{Updated miniboone
		neutrino oscillation results with increased data and new background studies},
	\href{https://doi.org/10.1103/PhysRevD.103.052002}{\emph{Phys. Rev. D}
		{\bfseries 103} (2021) 052002}.
	
	\bibitem{Mention:2011rk}
	G.~Mention, M.~Fechner, T.~Lasserre, T.A.~Mueller, D.~Lhuillier, M.~Cribier
	et~al., \emph{{The Reactor Antineutrino Anomaly}},
	\href{https://doi.org/10.1103/PhysRevD.83.073006}{\emph{Phys. Rev. D}
		{\bfseries 83} (2011) 073006} [\href{https://arxiv.org/abs/1101.2755}{{\tt
			arXiv:1101.2755}}].
	
	\bibitem{Mueller:2011nm}
	T.A.~Mueller et~al., \emph{{Improved Predictions of Reactor Antineutrino
			Spectra}}, \href{https://doi.org/10.1103/PhysRevC.83.054615}{\emph{Phys. Rev.
			C} {\bfseries 83} (2011) 054615} [\href{https://arxiv.org/abs/1101.2663}{{\tt
			arXiv:1101.2663}}].
	
	\bibitem{Huber:2011wv}
	P.~Huber, \emph{{On the determination of anti-neutrino spectra from nuclear
			reactors}}, \href{https://doi.org/10.1103/PhysRevC.85.029901}{\emph{Phys.
			Rev. C} {\bfseries 84} (2011) 024617}
	[\href{https://arxiv.org/abs/1106.0687}{{\tt arXiv:1106.0687}}].
	
	\bibitem{Collin:2016aqd}
	G.H.~Collin, C.A.~Arg\"uelles, J.M.~Conrad and M.H.~Shaevitz, \emph{{First
			Constraints on the Complete Neutrino Mixing Matrix with a Sterile Neutrino}},
	\href{https://doi.org/10.1103/PhysRevLett.117.221801}{\emph{Phys. Rev. Lett.}
		{\bfseries 117} (2016) 221801} [\href{https://arxiv.org/abs/1607.00011}{{\tt
			arXiv:1607.00011}}].
	
	\bibitem{Gariazzo:2017fdh}
	S.~Gariazzo, C.~Giunti, M.~Laveder and Y.F.~Li, \emph{{Updated Global 3+1
			Analysis of Short-BaseLine Neutrino Oscillations}},
	\href{https://doi.org/10.1007/JHEP06(2017)135}{\emph{JHEP} {\bfseries 06}
		(2017) 135} [\href{https://arxiv.org/abs/1703.00860}{{\tt
			arXiv:1703.00860}}].
	
	\bibitem{Esmaili:2013vza}
	A.~Esmaili and A.Y.~Smirnov, \emph{{Restricting the LSND and MiniBooNE sterile
			neutrinos with the IceCube atmospheric neutrino data}},
	\href{https://doi.org/10.1007/JHEP12(2013)014}{\emph{JHEP} {\bfseries 12}
		(2013) 014} [\href{https://arxiv.org/abs/1307.6824}{{\tt arXiv:1307.6824}}].
	
	\bibitem{Capozzi:2016vac}
	F.~Capozzi, C.~Giunti, M.~Laveder and A.~Palazzo, \emph{{Joint short- and
			long-baseline constraints on light sterile neutrinos}},
	\href{https://doi.org/10.1103/PhysRevD.95.033006}{\emph{Phys. Rev. D}
		{\bfseries 95} (2017) 033006} [\href{https://arxiv.org/abs/1612.07764}{{\tt
			arXiv:1612.07764}}].
	
	\bibitem{Dentler:2018sju}
	M.~Dentler, A.~Hern\'andez-Cabezudo, J.~Kopp, P.A.N.~Machado, M.~Maltoni,
	I.~Martinez-Soler et~al., \emph{{Updated Global Analysis of Neutrino
			Oscillations in the Presence of eV-Scale Sterile Neutrinos}},
	\href{https://doi.org/10.1007/JHEP08(2018)010}{\emph{JHEP} {\bfseries 08}
		(2018) 010} [\href{https://arxiv.org/abs/1803.10661}{{\tt
			arXiv:1803.10661}}].
	
	\bibitem{MicroBooNE:2021tya}
	{\scshape MicroBooNE} collaboration, \emph{{Search for an Excess of Electron
			Neutrino Interactions in MicroBooNE Using Multiple Final-State Topologies}},
	\href{https://doi.org/10.1103/PhysRevLett.128.241801}{\emph{Phys. Rev. Lett.}
		{\bfseries 128} (2022) 241801} [\href{https://arxiv.org/abs/2110.14054}{{\tt
			arXiv:2110.14054}}].
	
	\bibitem{MicroBooNE:2022wdf}
	{\scshape MicroBooNE} collaboration, \emph{{First constraints on light sterile
			neutrino oscillations from combined appearance and disappearance searches
			with the MicroBooNE detector}},  \href{https://arxiv.org/abs/2210.10216}{{\tt
			arXiv:2210.10216}}.
	
	\bibitem{Giunti:2021kab}
	C.~Giunti, Y.F.~Li, C.A.~Ternes and Z.~Xin, \emph{{Reactor antineutrino anomaly
			in light of recent flux model refinements}},
	\href{https://doi.org/10.1016/j.physletb.2022.137054}{\emph{Phys. Lett. B}
		{\bfseries 829} (2022) 137054} [\href{https://arxiv.org/abs/2110.06820}{{\tt
			arXiv:2110.06820}}].
	
	\bibitem{GALLEX:1994rym}
	{\scshape GALLEX} collaboration, \emph{{First results from the Cr-51 neutrino
			source experiment with the GALLEX detector}},
	\href{https://doi.org/10.1016/0370-2693(94)01586-2}{\emph{Phys. Lett. B}
		{\bfseries 342} (1995) 440}.
	
	\bibitem{GALLEX:1998kcz}
	{\scshape GALLEX} collaboration, \emph{{GALLEX solar neutrino observations:
			Results for GALLEX IV}},
	\href{https://doi.org/10.1016/S0370-2693(98)01579-2}{\emph{Phys. Lett. B}
		{\bfseries 447} (1999) 127}.
	
	\bibitem{Kaether:2010ag}
	F.~Kaether, W.~Hampel, G.~Heusser, J.~Kiko and T.~Kirsten, \emph{{Reanalysis of
			the GALLEX solar neutrino flux and source experiments}},
	\href{https://doi.org/10.1016/j.physletb.2010.01.030}{\emph{Phys. Lett. B}
		{\bfseries 685} (2010) 47} [\href{https://arxiv.org/abs/1001.2731}{{\tt
			arXiv:1001.2731}}].
	
	\bibitem{SAGE:1998fvr}
	{\scshape SAGE} collaboration, \emph{{Measurement of the response of the
			Russian-American gallium experiment to neutrinos from a Cr-51 source}},
	\href{https://doi.org/10.1103/PhysRevC.59.2246}{\emph{Phys. Rev. C}
		{\bfseries 59} (1999) 2246} [\href{https://arxiv.org/abs/hep-ph/9803418}{{\tt
			hep-ph/9803418}}].
	
	\bibitem{SAGE:1999uje}
	{\scshape SAGE} collaboration, \emph{{Measurement of the solar neutrino capture
			rate by SAGE and implications for neutrino oscillations in vacuum}},
	\href{https://doi.org/10.1103/PhysRevLett.83.4686}{\emph{Phys. Rev. Lett.}
		{\bfseries 83} (1999) 4686}
	[\href{https://arxiv.org/abs/astro-ph/9907131}{{\tt astro-ph/9907131}}].
	
	\bibitem{Abdurashitov:2005tb}
	J.N.~Abdurashitov et~al., \emph{{Measurement of the response of a Ga solar
			neutrino experiment to neutrinos from an Ar-37 source}},
	\href{https://doi.org/10.1103/PhysRevC.73.045805}{\emph{Phys. Rev. C}
		{\bfseries 73} (2006) 045805}
	[\href{https://arxiv.org/abs/nucl-ex/0512041}{{\tt nucl-ex/0512041}}].
	
	\bibitem{SAGE:2009eeu}
	{\scshape SAGE} collaboration, \emph{{Measurement of the solar neutrino capture
			rate with gallium metal. III: Results for the 2002--2007 data-taking
			period}}, \href{https://doi.org/10.1103/PhysRevC.80.015807}{\emph{Phys. Rev.
			C} {\bfseries 80} (2009) 015807} [\href{https://arxiv.org/abs/0901.2200}{{\tt
			arXiv:0901.2200}}].
	
	\bibitem{Barinov:2021asz}
	V.V.~Barinov et~al., \emph{{Results from the Baksan Experiment on Sterile
			Transitions (BEST)}},
	\href{https://doi.org/10.1103/PhysRevLett.128.232501}{\emph{Phys. Rev. Lett.}
		{\bfseries 128} (2022) 232501} [\href{https://arxiv.org/abs/2109.11482}{{\tt
			arXiv:2109.11482}}].
	
	\bibitem{Barinov:2022wfh}
	V.V.~Barinov et~al., \emph{{Search for electron-neutrino transitions to sterile
			states in the BEST experiment}},
	\href{https://doi.org/10.1103/PhysRevC.105.065502}{\emph{Phys. Rev. C}
		{\bfseries 105} (2022) 065502} [\href{https://arxiv.org/abs/2201.07364}{{\tt
			arXiv:2201.07364}}].
	
	\bibitem{Giunti:2022btk}
	C.~Giunti, Y.F.~Li, C.A.~Ternes, O.~Tyagi and Z.~Xin, \emph{{Gallium Anomaly:
			Critical View from the Global Picture of $\nu_{e}$ and $\bar\nu_{e}$
			Disappearance}},  \href{https://arxiv.org/abs/2209.00916}{{\tt
			arXiv:2209.00916}}.
	
	\bibitem{Arguelles:2022bvt}
	C.A.~Arg\"uelles, T.~Bert\'olez-Mart\'\i{}nez and J.~Salvado, \emph{{Impact of
			Wave Packet Separation in Low-Energy Sterile Neutrino Searches}},
	\href{https://arxiv.org/abs/2201.05108}{{\tt arXiv:2201.05108}}.
	
	\bibitem{Asaka:2005an}
	T.~Asaka, S.~Blanchet and M.~Shaposhnikov, \emph{{The nuMSM, dark matter and
			neutrino masses}},
	\href{https://doi.org/10.1016/j.physletb.2005.09.070}{\emph{Phys. Lett. B}
		{\bfseries 631} (2005) 151} [\href{https://arxiv.org/abs/hep-ph/0503065}{{\tt
			hep-ph/0503065}}].
	
	\bibitem{Asaka:2005pn}
	T.~Asaka and M.~Shaposhnikov, \emph{{The $\nu$MSM, dark matter and baryon
			asymmetry of the universe}},
	\href{https://doi.org/10.1016/j.physletb.2005.06.020}{\emph{Phys. Lett. B}
		{\bfseries 620} (2005) 17} [\href{https://arxiv.org/abs/hep-ph/0505013}{{\tt
			hep-ph/0505013}}].
	
	\bibitem{Boyarsky:2009ix}
	A.~Boyarsky, O.~Ruchayskiy and M.~Shaposhnikov, \emph{{The Role of sterile
			neutrinos in cosmology and astrophysics}},
	\href{https://doi.org/10.1146/annurev.nucl.010909.083654}{\emph{Ann. Rev.
			Nucl. Part. Sci.} {\bfseries 59} (2009) 191}
	[\href{https://arxiv.org/abs/0901.0011}{{\tt arXiv:0901.0011}}].
	
	\bibitem{Adhikari:2016bei}
	M.~Drewes et~al., \emph{{A White Paper on keV Sterile Neutrino Dark Matter}},
	\href{https://doi.org/10.1088/1475-7516/2017/01/025}{\emph{JCAP} {\bfseries
			01} (2017) 025} [\href{https://arxiv.org/abs/1602.04816}{{\tt
			arXiv:1602.04816}}].
	
	\bibitem{VIOLLIER199379}
	R.~Viollier, D.~Trautmann and G.~Tupper, \emph{Supermassive neutrino stars and
		galactic nuclei},
	\href{https://doi.org/https://doi.org/10.1016/0370-2693(93)91141-9}{\emph{Physics
			Letters B} {\bfseries 306} (1993) 79}.
	
	\bibitem{Bilic:2001iv}
	N.~Bilic, R.J.~Lindebaum, G.B.~Tupper and R.D.~Viollier, \emph{{On the
			formation of degenerate heavy neutrino stars}},
	\href{https://doi.org/10.1016/S0370-2693(01)00836-X}{\emph{Phys. Lett. B}
		{\bfseries 515} (2001) 105}
	[\href{https://arxiv.org/abs/astro-ph/0106209}{{\tt astro-ph/0106209}}].
	
	\bibitem{deHolanda:2003tx}
	P.C.~de~Holanda and A.Y.~Smirnov, \emph{{Homestake result, sterile neutrinos
			and low-energy solar neutrino experiments}},
	\href{https://doi.org/10.1103/PhysRevD.69.113002}{\emph{Phys. Rev. D}
		{\bfseries 69} (2004) 113002}
	[\href{https://arxiv.org/abs/hep-ph/0307266}{{\tt hep-ph/0307266}}].
	
	\bibitem{deHolanda:2010am}
	P.C.~de~Holanda and A.Y.~Smirnov, \emph{{Solar neutrino spectrum, sterile
			neutrinos and additional radiation in the Universe}},
	\href{https://doi.org/10.1103/PhysRevD.83.113011}{\emph{Phys. Rev. D}
		{\bfseries 83} (2011) 113011} [\href{https://arxiv.org/abs/1012.5627}{{\tt
			arXiv:1012.5627}}].
	
	\bibitem{Dev:2012bd}
	P.S.~Bhupal~Dev and A.~Pilaftsis, \emph{{Light and Superlight Sterile Neutrinos
			in the Minimal Radiative Inverse Seesaw Model}},
	\href{https://doi.org/10.1103/PhysRevD.87.053007}{\emph{Phys. Rev. D}
		{\bfseries 87} (2013) 053007} [\href{https://arxiv.org/abs/1212.3808}{{\tt
			arXiv:1212.3808}}].
	
	\bibitem{Liao:2014ola}
	W.~Liao, Y.~Luo and X.-H.~Wu, \emph{{Effect of interaction with neutrons in
			matter on flavor conversion of super-light sterile neutrino with active
			neutrino}}, \href{https://doi.org/10.1007/JHEP06(2014)069}{\emph{JHEP}
		{\bfseries 06} (2014) 069} [\href{https://arxiv.org/abs/1403.2559}{{\tt
			arXiv:1403.2559}}].
	
	\bibitem{Divari:2016jos}
	P.C.~Divari and J.D.~Vergados, \emph{{Neutrino oscillations in the presence of
			super-light sterile neutrinos}},
	\href{https://doi.org/10.1142/S0217751X16501232}{\emph{Int. J. Mod. Phys. A}
		{\bfseries 31} (2016) 1650123} [\href{https://arxiv.org/abs/1602.08690}{{\tt
			arXiv:1602.08690}}].
	
	\bibitem{Abe:2016nxk}
	{\scshape Super-Kamiokande} collaboration, \emph{{Solar Neutrino Measurements
			in Super-Kamiokande-IV}},
	\href{https://doi.org/10.1103/PhysRevD.94.052010}{\emph{Phys. Rev. D}
		{\bfseries 94} (2016) 052010} [\href{https://arxiv.org/abs/1606.07538}{{\tt
			arXiv:1606.07538}}].
	
	\bibitem{Aharmim:2011vm}
	{\scshape SNO} collaboration, \emph{{Combined Analysis of all Three Phases of
			Solar Neutrino Data from the Sudbury Neutrino Observatory}},
	\href{https://doi.org/10.1103/PhysRevC.88.025501}{\emph{Phys. Rev. C}
		{\bfseries 88} (2013) 025501} [\href{https://arxiv.org/abs/1109.0763}{{\tt
			arXiv:1109.0763}}].
	
	\bibitem{Agostini:2017ixy}
	{\scshape Borexino} collaboration, \emph{{First Simultaneous Precision
			Spectroscopy of $pp$, $^7$Be, and $pep$ Solar Neutrinos with Borexino
			Phase-II}}, \href{https://doi.org/10.1103/PhysRevD.100.082004}{\emph{Phys.
			Rev. D} {\bfseries 100} (2019) 082004}
	[\href{https://arxiv.org/abs/1707.09279}{{\tt arXiv:1707.09279}}].
	
	\bibitem{deGouvea:2022kma}
	A.~de~Gouv\^ea, G.~Jusino~S\'anchez and K.J.~Kelly, \emph{{Very light sterile
			neutrinos at NOvA and T2K}},
	\href{https://doi.org/10.1103/PhysRevD.106.055025}{\emph{Phys. Rev. D}
		{\bfseries 106} (2022) 055025} [\href{https://arxiv.org/abs/2204.09130}{{\tt
			arXiv:2204.09130}}].
	
	\bibitem{Kelly:2020fkv}
	K.J.~Kelly, P.A.N.~Machado, S.J.~Parke, Y.F.~Perez-Gonzalez and R.Z.~Funchal,
	\emph{{Neutrino mass ordering in light of recent data}},
	\href{https://doi.org/10.1103/PhysRevD.103.013004}{\emph{Phys. Rev. D}
		{\bfseries 103} (2021) 013004} [\href{https://arxiv.org/abs/2007.08526}{{\tt
			arXiv:2007.08526}}].
	
	\bibitem{Acero:2022wqg}
	M.A.~Acero et~al., \emph{{White Paper on Light Sterile Neutrino Searches and
			Related Phenomenology}},  \href{https://arxiv.org/abs/2203.07323}{{\tt
			arXiv:2203.07323}}.
	
	\bibitem{Wong:2011ip}
	Y.Y.Y.~Wong, \emph{{Neutrino mass in cosmology: status and prospects}},
	\href{https://doi.org/10.1146/annurev-nucl-102010-130252}{\emph{Ann. Rev.
			Nucl. Part. Sci.} {\bfseries 61} (2011) 69}
	[\href{https://arxiv.org/abs/1111.1436}{{\tt arXiv:1111.1436}}].
	
	\bibitem{FrancoAbellan:2021hdb}
	G.~Franco~Abell\'an, Z.~Chacko, A.~Dev, P.~Du, V.~Poulin and Y.~Tsai,
	\emph{{Improved cosmological constraints on the neutrino mass and lifetime}},
	\href{https://doi.org/10.1007/JHEP08(2022)076}{\emph{JHEP} {\bfseries 08}
		(2022) 076} [\href{https://arxiv.org/abs/2112.13862}{{\tt
			arXiv:2112.13862}}].
	
	\bibitem{Thakore:2018lgn}
	T.~Thakore, M.M.~Devi, S.~Kumar~Agarwalla and A.~Dighe, \emph{{Active-sterile
			neutrino oscillations at INO-ICAL over a wide mass-squared range}},
	\href{https://doi.org/10.1007/JHEP08(2018)022}{\emph{JHEP} {\bfseries 08}
		(2018) 022} [\href{https://arxiv.org/abs/1804.09613}{{\tt
			arXiv:1804.09613}}].
	
	\bibitem{Kumar:2017sdq}
	{\scshape ICAL} collaboration, \emph{{Physics Potential of the ICAL detector at
			the India-based Neutrino Observatory (INO)}},
	\href{https://doi.org/10.1007/s12043-017-1373-4}{\emph{Pramana} {\bfseries
			88} (2017) 79} [\href{https://arxiv.org/abs/1505.07380}{{\tt
			arXiv:1505.07380}}].
	
	\bibitem{Gandhi:2015xza}
	R.~Gandhi, B.~Kayser, M.~Masud and S.~Prakash, \emph{{The impact of sterile
			neutrinos on CP measurements at long baselines}},
	\href{https://doi.org/10.1007/JHEP11(2015)039}{\emph{JHEP} {\bfseries 11}
		(2015) 039} [\href{https://arxiv.org/abs/1508.06275}{{\tt
			arXiv:1508.06275}}].
	
	\bibitem{Choubey:2017ppj}
	S.~Choubey, D.~Dutta and D.~Pramanik, \emph{{Measuring the Sterile Neutrino CP
			Phase at DUNE and T2HK}},
	\href{https://doi.org/10.1140/epjc/s10052-018-5816-y}{\emph{Eur. Phys. J. C}
		{\bfseries 78} (2018) 339} [\href{https://arxiv.org/abs/1711.07464}{{\tt
			arXiv:1711.07464}}].
	
	\bibitem{Dighe:2007uf}
	A.~Dighe and S.~Ray, \emph{{Signatures of heavy sterile neutrinos at long
			baseline experiments}},
	\href{https://doi.org/10.1103/PhysRevD.76.113001}{\emph{Phys. Rev. D}
		{\bfseries 76} (2007) 113001} [\href{https://arxiv.org/abs/0709.0383}{{\tt
			arXiv:0709.0383}}].
	
	\bibitem{Ray:2010tea}
	S.~Ray, \emph{{Neutrino oscillation phenomenology with fermions beyond the
			standard model}}, Ph.D. thesis, Tata Inst., 2010.
	
	\bibitem{Klop:2014ima}
	N.~Klop and A.~Palazzo, \emph{{Imprints of CP violation induced by sterile
			neutrinos in T2K data}},
	\href{https://doi.org/10.1103/PhysRevD.91.073017}{\emph{Phys. Rev. D}
		{\bfseries 91} (2015) 073017} [\href{https://arxiv.org/abs/1412.7524}{{\tt
			arXiv:1412.7524}}].
	
	\bibitem{Agarwalla:2016xxa}
	S.K.~Agarwalla, S.S.~Chatterjee and A.~Palazzo, \emph{{Physics Reach of DUNE
			with a Light Sterile Neutrino}},
	\href{https://doi.org/10.1007/JHEP09(2016)016}{\emph{JHEP} {\bfseries 09}
		(2016) 016} [\href{https://arxiv.org/abs/1603.03759}{{\tt
			arXiv:1603.03759}}].
	
	\bibitem{Haba:2018klh}
	N.~Haba, Y.~Mimura and T.~Yamada, \emph{{$\theta_{23}$ octant measurement in
			$3+1$ neutrino oscillations in T2HKK}},
	\href{https://doi.org/10.1103/PhysRevD.101.075034}{\emph{Phys. Rev. D}
		{\bfseries 101} (2020) 075034} [\href{https://arxiv.org/abs/1812.10940}{{\tt
			arXiv:1812.10940}}].
	
	\bibitem{Sharma:2022qeo}
	K.~Sharma and S.~Patra, \emph{{Study of matter effects in the presence of
			sterile neutrino using OMSD approximation}},
	\href{https://arxiv.org/abs/2207.03249}{{\tt arXiv:2207.03249}}.
	
	\bibitem{Kamo:2002sj}
	Y.~Kamo, S.~Yajima, Y.~Higasida, S.-I.~Kubota, S.~Tokuo and J.-I.~Ichihara,
	\emph{{Analytical calculations of four neutrino oscillations in matter}},
	\href{https://doi.org/10.1140/epjc/s2003-01138-0}{\emph{Eur. Phys. J. C}
		{\bfseries 28} (2003) 211} [\href{https://arxiv.org/abs/hep-ph/0209097}{{\tt
			hep-ph/0209097}}].
	
	\bibitem{Li:2018ezt}
	W.~Li, J.~Ling, F.~Xu and B.~Yue, \emph{{Matter Effect of Light Sterile
			Neutrino: An Exact Analytical Approach}},
	\href{https://doi.org/10.1007/JHEP10(2018)021}{\emph{JHEP} {\bfseries 10}
		(2018) 021} [\href{https://arxiv.org/abs/1808.03985}{{\tt
			arXiv:1808.03985}}].
	
	\bibitem{Parke:2019jyu}
	S.J.~Parke and X.~Zhang, \emph{{Compact Perturbative Expressions for
			Oscillations with Sterile Neutrinos in Matter}},
	\href{https://doi.org/10.1103/PhysRevD.101.056005}{\emph{Phys. Rev. D}
		{\bfseries 101} (2020) 056005} [\href{https://arxiv.org/abs/1905.01356}{{\tt
			arXiv:1905.01356}}].
	
	\bibitem{Yue:2019qat}
	B.~Yue, W.~Li, J.~Ling and F.~Xu, \emph{{A compact analytical approximation for
			a light sterile neutrino oscillation in matter}},
	\href{https://doi.org/10.1088/1674-1137/33/S2/005}{\emph{Chin. Phys. C}
		{\bfseries 44} (2020) 103001} [\href{https://arxiv.org/abs/1906.03781}{{\tt
			arXiv:1906.03781}}].
	
	\bibitem{Reyimuaji:2019wbn}
	Y.~Reyimuaji and C.~Liu, \emph{{Prospects of light sterile neutrino searches in
			long-baseline neutrino oscillations}},
	\href{https://doi.org/10.1007/JHEP06(2020)094}{\emph{JHEP} {\bfseries 06}
		(2020) 094} [\href{https://arxiv.org/abs/1911.12524}{{\tt
			arXiv:1911.12524}}].
	
	\bibitem{Fong:2022oim}
	C.S.~Fong, \emph{{Analytic Neutrino Oscillation Probabilities}},
	\href{https://arxiv.org/abs/2210.09436}{{\tt arXiv:2210.09436}}.
	
	\bibitem{DUNE:2015lol}
	{\scshape DUNE} collaboration, \emph{{Long-Baseline Neutrino Facility (LBNF)
			and Deep Underground Neutrino Experiment (DUNE)}: {Conceptual Design Report,
			Volume 2: The Physics Program for DUNE at LBNF}},
	\href{https://arxiv.org/abs/1512.06148}{{\tt arXiv:1512.06148}}.
	
	\bibitem{Lindner:2001fx}
	M.~Lindner, T.~Ohlsson and W.~Winter, \emph{{A Combined treatment of neutrino
			decay and neutrino oscillations}},
	\href{https://doi.org/10.1016/S0550-3213(01)00237-1}{\emph{Nucl. Phys. B}
		{\bfseries 607} (2001) 326} [\href{https://arxiv.org/abs/hep-ph/0103170}{{\tt
			hep-ph/0103170}}].
	
	\bibitem{Akhmedov:2004ny}
	E.K.~Akhmedov, R.~Johansson, M.~Lindner, T.~Ohlsson and T.~Schwetz,
	\emph{{Series expansions for three flavor neutrino oscillation probabilities
			in matter}}, \href{https://doi.org/10.1088/1126-6708/2004/04/078}{\emph{JHEP}
		{\bfseries 04} (2004) 078} [\href{https://arxiv.org/abs/hep-ph/0402175}{{\tt
			hep-ph/0402175}}].
	
	\bibitem{Huber:2004ka}
	P.~Huber, M.~Lindner and W.~Winter, \emph{{Simulation of long-baseline neutrino
			oscillation experiments with GLoBES (General Long Baseline Experiment
			Simulator)}}, \href{https://doi.org/10.1016/j.cpc.2005.01.003}{\emph{Comput.
			Phys. Commun.} {\bfseries 167} (2005) 195}
	[\href{https://arxiv.org/abs/hep-ph/0407333}{{\tt hep-ph/0407333}}].
	
	\bibitem{Huber:2007ji}
	P.~Huber, J.~Kopp, M.~Lindner, M.~Rolinec and W.~Winter, \emph{{New features in
			the simulation of neutrino oscillation experiments with GLoBES 3.0: General
			Long Baseline Experiment Simulator}},
	\href{https://doi.org/10.1016/j.cpc.2007.05.004}{\emph{Comput. Phys. Commun.}
		{\bfseries 177} (2007) 432} [\href{https://arxiv.org/abs/hep-ph/0701187}{{\tt
			hep-ph/0701187}}].
	
	\bibitem{DUNE:2020fgq}
	{\scshape DUNE} collaboration, \emph{{Prospects for beyond the Standard Model
			physics searches at the Deep Underground Neutrino Experiment}},
	\href{https://doi.org/10.1140/epjc/s10052-021-09007-w}{\emph{Eur. Phys. J. C}
		{\bfseries 81} (2021) 322} [\href{https://arxiv.org/abs/2008.12769}{{\tt
			arXiv:2008.12769}}].
	
	\bibitem{MINOS:2017cae}
	{\scshape MINOS+} collaboration, \emph{{Search for sterile neutrinos in MINOS
			and MINOS+ using a two-detector fit}},
	\href{https://doi.org/10.1103/PhysRevLett.122.091803}{\emph{Phys. Rev. Lett.}
		{\bfseries 122} (2019) 091803} [\href{https://arxiv.org/abs/1710.06488}{{\tt
			arXiv:1710.06488}}].
	
\end{thebibliography}
\end{document}